\documentclass[useAMS,usenatbib]{mn2e}
\usepackage{amssymb,amsmath,epsfig,times, natbib}

\defcitealias{Simionescu2011}{Simionescu et al. (2011a)}
\defcitealias{Simionescu2011conf}{Simionescu et al. (2011b)}
\defcitealias{Eckert2011a}{Eckert et al. (2011a)}
\defcitealias{Eckert2011b}{Eckert et al. (2011b)}

\title[PKS 0745-191 at the virial radius]{Further X-ray observations
  of the galaxy cluster PKS 0745-191 to the virial radius and beyond}
\author[S. A. Walker et al.]{S. A. Walker,$^1$\thanks{Email: 
    swalker@ast.cam.ac.uk} A. C. Fabian,$^1$ J. S. Sanders$^1$, M. R. George$^2$ \\
  $^1$Institute of Astronomy, Madingley Road, Cambridge CB3 0HA \\
  $^2$Department of Astronomy, University of California, Berkeley, CA 94720, USA\\
    \\
   \\
   \\
}
\date{}

\begin{document}

\maketitle

\begin{abstract}
 We use new \emph{Suzaku} observations of PKS 0745-191 to measure the thermodynamic properties of its ICM out to and beyond r$_{200}$ (reaching 1.25r$_{200}$) with better accuracy than previously achieved, owing to a more accurate and better understood background model. We investigate and resolve the tensions between the previous Suzaku and \emph{ROSAT} results for PKS 0745-191, which are found to be principally caused by incorrect background modelling in the previous \emph{Suzaku} analysis. We investigate in depth the systematic errors affecting this observation, and present temperature, density, entropy and gas mass fraction profiles reaching out to and beyond the virial radius. We find that the entropy profile flattens in the outskirts as originally observed in the previous \emph{Suzaku} analysis, but that the flattening starts at larger radius. The flattening of the entropy profile and our mass analysis suggests that outside $\sim$17$'$ ($\sim$1.9 Mpc) the ICM is out of hydrostatic equilibrium or the presence of significant non-thermal pressure support. 
\end{abstract}


\section{Introduction}

Accurate measurements of the intracluster medium (ICM) of galaxy clusters to the virial radius (taken to be r$_{200}$\footnote{r$_{200}$ is the radius within which the mean density of the cluster is 200 times the critical density required for a flat universe, $\rho_{c}$. The mass enclosed within r$_{200}$ is M$_{200}$=4/3$\pi\rho_{c}$r$_{200}^{3}$. }), are important for many reasons. The virial radius represents the boundary within which the cluster ICM is expected to be in hydrostatic equilibrium, and outside of which matter is still accreting onto the cluster as it continues to form. Studying cluster outskirts allows the cluster formation process to be better understood, thus constraining simulations of cluster formation (such as \citealt{Roncarelli2006}, \citealt{Burns2010}) while allowing us to constrain models for the baryon fractions of clusters (for example \citealt{Young2011}) which are dependent on non-gravitational processes such as feedback and cooling.

We can also investigate where the assumption of hydrostatic equilibrium breaks down. The hydrostatic equilibrium assumption is important because it is used when calculating the masses of galaxy clusters, as we see later in section \ref{massanalysis}. Accurate cluster masses are required for using galaxy clusters as probes of cosmological parameters using the cluster mass function \citep{Vikhlinin2009}, and understanding the gas mass fraction to the virial radius improves the reliability with which it can be used as a cosmological probe \citep{Allen2008}.

In \citet{George2009} (hereafter G09) the ICM of the galaxy cluster PKS 0745-191 was investigated to beyond the virial radius they calculated for the first time. This exploited \emph{Suzaku's} low instrumental background, resulting from its low orbit which places it inside the Earth's magnetopause, allowing the low surface brightness outskirts to be probed. This development was made possible by the fact that PKS 0745-191 is very X-ray bright (it is the brightest cluster beyond z $=$ 0.1).

One of the most interesting findings of G09 was that the entropy profile of the ICM flattened in the outskirts outside 10$'$=1.1Mpc, deviating significantly from the theoretical prediction of a r$^{1.1}$ power law increase assuming gravitational structure formation \citep{Voit2005}. However the location of PKS 0745-191 just 3 degrees above the galactic midplane means that it has a complicated and spatially variable background and absorbing column density which could not be properly understood due to the lack of suitable background pointings when G09 was written. In order to fully understand the properties of PKS 0745-191 in the outskirts, an accurate background model is needed whose errors are thoroughly understood. 

Further clusters have since been studied to the virial radius with Suzaku [Abell 2204 \citep{Reiprich2009}; Abell 1975 \citep{Bautz2009}; Abell 1413 \citep{Hoshino2010}; Abell 1689 \citep{Kawaharada2010}; the Perseus cluster \citepalias{Simionescu2011}; Abell 2142 \citep{Akamatsu2011}; Abell 2029 \citep{Walker2012_A2029} and Hydra A \citep{Sato2012}], and a statistical sample with general trends has started to emerge. One common feature that is observed for most of these clusters is the flattening of the entropy profile from around 0.5r$_{200}$ outwards, causing the entropy to be below the powerlaw prediction of \citet{Voit2005} in the outskirts. One explanation of this is that the cluster ICM is out of hydrostatic equilibrium in the outskirts, and that non-thermal pressure support from turbulence and bulk motions is important, as is found in the numerical simulations of \citet{Burns2010}. However the results for the Perseus cluster (which because of its closeness and thus large angular extent has been studied with better resolution than the other clusters) suggest that the entropy profile flattening is a consequence of the gas density in the outskirts being overestimated due to gas clumping, which also causes the measurements of the gas mass fraction and gas pressure in the outskirts to be higher than expected.   

Numerical simulations of galaxy cluster outskirts indicate that the contributions of non-thermal pressure support increase with radius in relaxed clusters, leading to an underestimate 
of the hydrostatic mass if only the thermal pressure (which is measured directly in X-ray observations) is used in the hydrostatic equilibrium equation. Random
gas motions become increasing important in the outskirts of galaxy cluster (from r$_{500}$ outwards) due to merging activity, the continual accretion of gas onto the cluster along large
scale structure filaments, and due to supersonic motions of galaxies through the ICM.   The simulations of \citet{Lau2009}
suggested that the non-thermal pressure contribution in relaxed clusters increases with radius to around 20-30 percent at r$_{200}$, resulting in the hydrostatic mass derived from the thermal
pressure underestimating the actual mass by $\sim$10 percent at r$_{500}$, rising to $\sim$ 20 percent at r$_{200}$. Similar results for the mass underestimate in the cluster outskirts were 
found in the simulations of \citet{Burns2010}, who found that at the thermal energy density decreases with radius while the energy density in turbulent kinetic energy increases with radius, with 
both contributions being approximately equal at $\sim$40 percent each at r$_{200}$, with the remaining $\sim$20 percent being provided by radial kinetic energy due  to infalling gas. 

Simulations also indicate that gas clumping plays an important role in effecting the measured gas density in the outskirts (\citealt{Mathiesen1999}, \citealt{Nagai2011}). This occurs because the ICM emits X-rays due to thermal 
bremsstrahlung, with the flux being proportional n$_{H}^{2}$, so if the ICM is clumped then this leads to more X-ray emission than if it is uniform with the same average density because
$\langle$n$_{H}^{2}\rangle$ $>$ $\langle$n$_{H}\rangle^{2}$. Therefore, if the ICM is assumed to be unform when it is in reality clumped, the gas density will be overestimated, leading
 to an overestimate of the gas mass fraction and the gas pressure, and an underestimate of the gas entropy (S=kT/n$_{e}^{2/3}$). In \citet{Nagai2011} gas clumping is found to 
be most important outside r$_{200}$, though the scatter between different simulated clusters is large. The exact nature of the clumping (its size scale and distribution, its azimuthal variation
and relation to large scale structure filaments) remains uncertain both in simulations (where it depends sensitively on the precise ICM physics, such as the rate of cooling 
and star formation) and in observation.

Tensions between the \emph{Suzaku} results of G09 and measurements with other X-ray observatories have recently been highlighted in \citetalias{Eckert2011a} (hereafter E11a), which used \emph{ROSAT} PSPC observations, and found density and surface brightness profiles which are inconsistent with those presented in G09 in the outskirts. In addition the temperature profile found in G09 is much lower than that obtained in the 700-1500 kpc region using \emph{XMM-Newton} \citep{Snowden2008} and \emph{Swift} \citep{Moretti2011}, and the mass determination in G09 is lower than and inconsistent with previous measurements (see table \ref{compare_masses} for a comparison). In addition, \citet{Sato2012} has shown that while the entropy profiles (scaled by the average ICM temperature) for the clusters investigated to r$_{200}$ with \emph{Suzaku} so far are consistent with one other with low scatter, the entropy profile found in G09 for PKS 0745-191 is lower in the outskirts and inconsistent with the other profiles.  

E11a attributed these tensions to incorrect background modelling. As no nearby background pointings using Suzaku were available when
G09 was written, the background was modelled using data from the Lockman Hole, which was found from Rosat All Sky Survey (RASS) data to have a similar count rate in the 1.0-2.0 keV band as the regions surrounding PKS 0745-191. The Lockman Hole data were also chosen because they were taken at a similar time as the original PKS 0745-191 data, thus minimising the effects of time variability of the contamination on the optical blocking filter (OBF) in front of the detectors (which was not completely understood when G09 was written). However the Lockman Hole is located at a much higher galactic latitude than PKS 0745-191, meaning that the galactic column density is much lower, and that the source of the 1-2 keV emission observed in the Lockman Hole data would not be expected to penetrate the absorbing column at the latitude of PKS 0745-191. Another contribution to the background at low galactic latitude must exist to compensate for the decrease in flux resulting from greater absorption, and its spectrum may be significantly different.   

To further study the properties of the cluster outskirts and to better
constrain our model of the background flux, we have obtained new \emph{Suzaku}
observations near PKS 0745-191. In this paper we analyse these new data along
with the original data used in G09 and other archival observations to
investigate the source of inconsistency seen in previous work.

\begin{figure*}
  \begin{center}
    \leavevmode
      \epsfig{figure=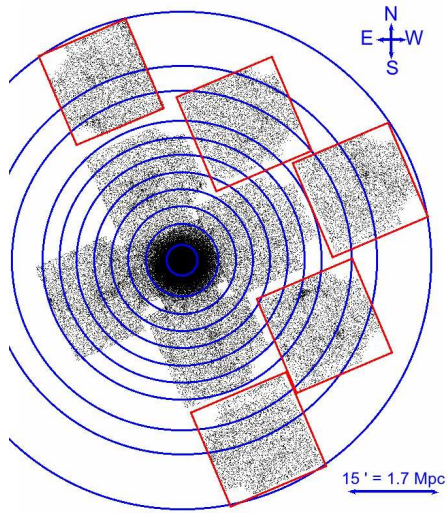,
        width=0.35\linewidth}
         \epsfig{figure=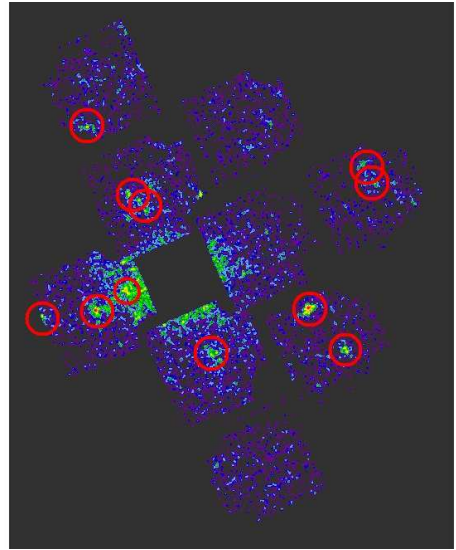,
        width=0.45\linewidth}     
        
      \caption{\emph{Left}:Mosaic of Suzaku pointings. Pointings used in G09 are in the central cross, and the new pointings extending the cross are shown in red. Annuli are between 0.0$'$-2.5$'$, 2.5$'$-6.0$'$, 6.0$'$-8.8$'$, 8.8$'$-11.7$'$, 11.7$'$-14.5$'$, 14.5$'$-17.3$'$, 17.3$'$-20.2$'$, 20.2$'$-23$'$, 23$'$-28$'$, 28$'$-32$'$ and 32$'$-41$'$. \emph{Right}: Point sources removed from the analysis using a threshold flux of 10$^{-13}$ erg cm$^{-2}$ s$^{-1}$ in the 2-10 keV band. In both panels the vignetting has not been corrected for, and the images are for illustrative purposes only.}
      \label{annuliPKS}
  \end{center}
\end{figure*}

\begin{figure*}
  \begin{center}
    \leavevmode
    \hbox{
      \epsfig{figure=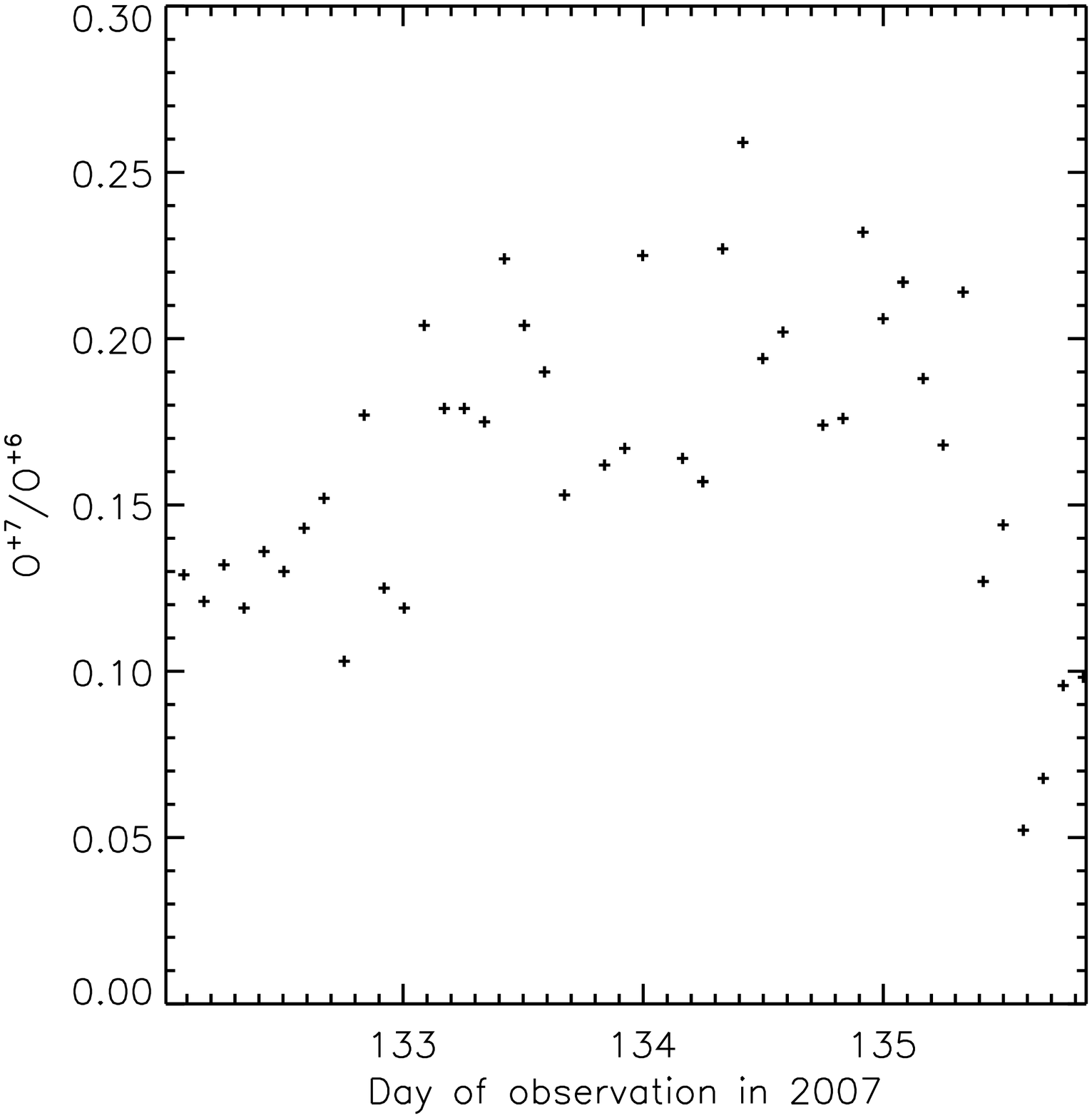,
        width=0.31\linewidth}
         \epsfig{figure=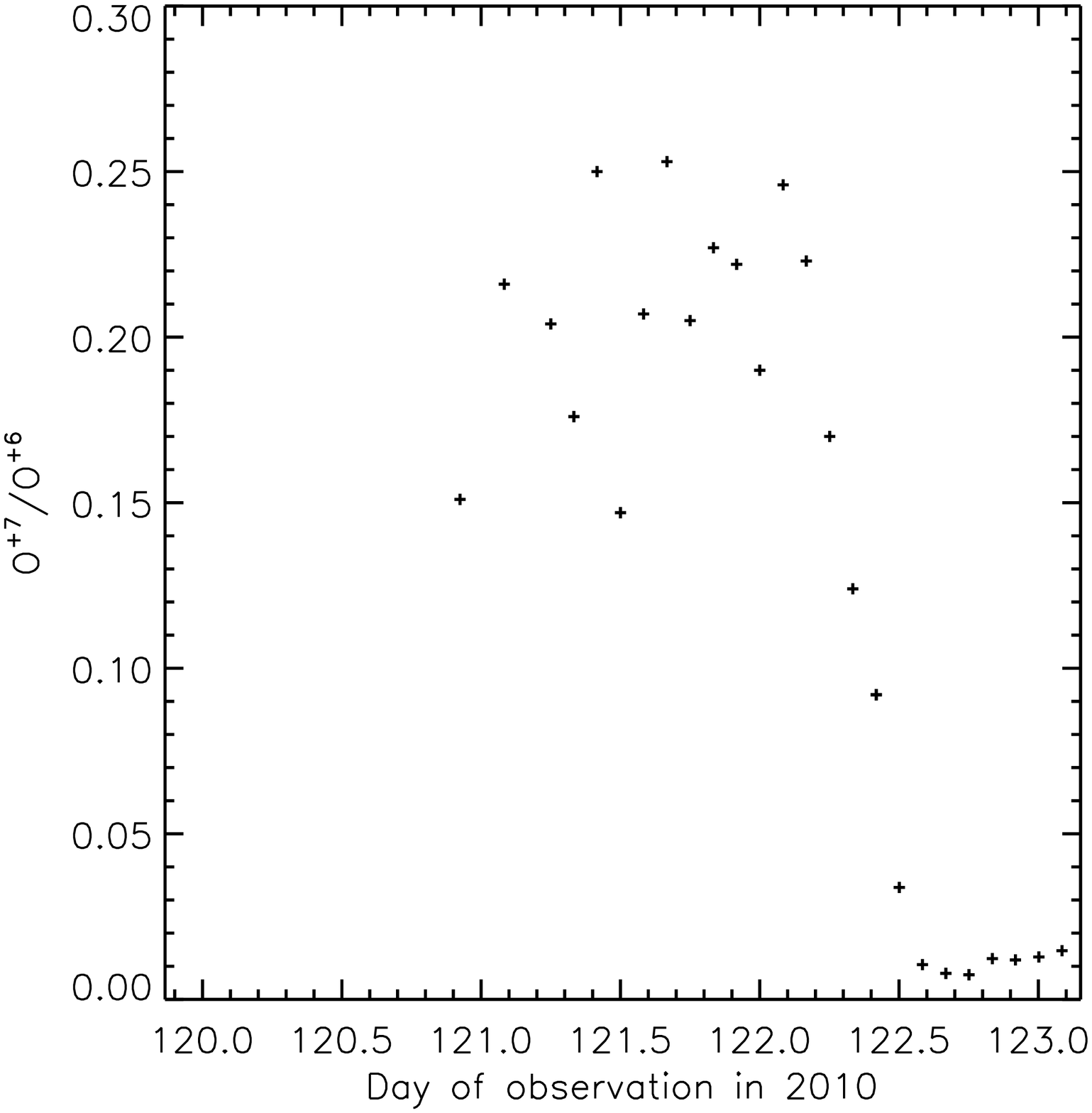,
        width=0.31\linewidth}   
         \epsfig{figure=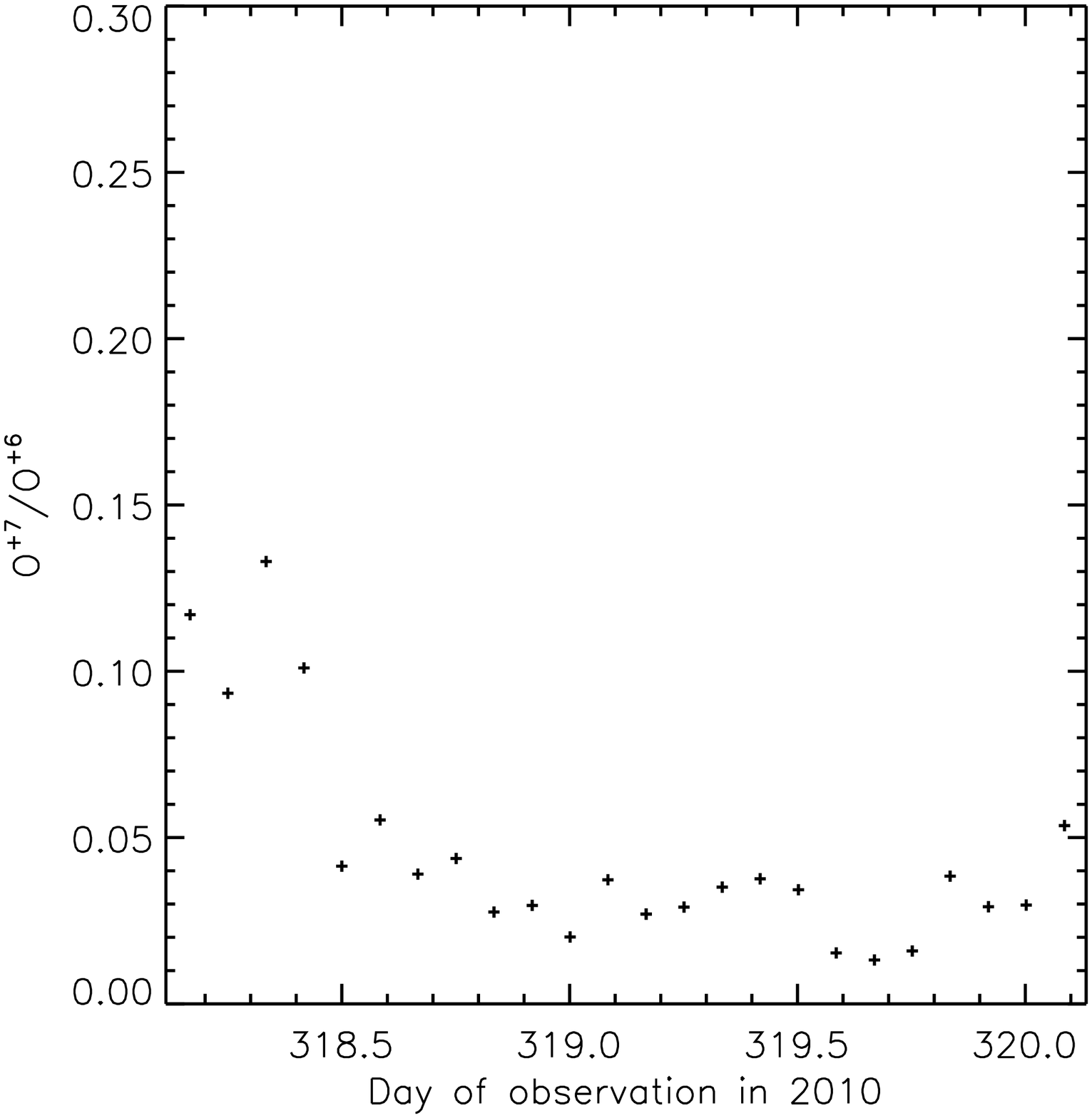,
        width=0.31\linewidth} 
        }
      \caption{Checking for SWCX contamination during the periods of observation using the  O$^{+7}$/O$^{+6}$ ratio obtained with the ACE spacecraft. Contamination is important when O$^{+7}$/O$^{+6}$ is greater then $\sim$ 0.2.  \emph{Left}: Original data used in \citep{George2009} \emph{Centre}: Pointings taken in May 2010. \emph{Right}: Pointings taken in November 2010.  }
      \label{ACE}
  \end{center}
\end{figure*}

\section{Observations and Data Reduction}

To supplement the original 5 pointings in the cross formation used in G09, which were taken in 2007 May, five more pointings were taken in 2010 May and 2010 November, which extend the cross to the North, West and South and fill in the corners between these three arms, as shown in Fig. \ref{annuliPKS}. This allows a local background determination to be made and allows spatial variations in the background parameters to be better understood, which is important for quantifying the uncertainty in the background and its affect on the cluster properties derived from spectral fits. 

The data were reduced using the same method as that described in \citet{Walker2012_A2029}. The ten \emph{Suzaku} XIS (X-ray Imaging Spectrometer) pointings of PKS 0745-191 were reduced using the ftool \textsc{aepipeline} which performs the standard filtering described in section 6.2 of the Suzaku ABC guide\footnote{heasarc.nasa.gov/docs/suzaku/analysis/abc/} using the latest calibrations in HEASoft version 6.11.1. The screening criterion COR$>$6 (which removes time intervals where the cut off rigidity was below 6 GeV) was included which reduces the good time interval (GTI) total exposure time to the values shown in Table \ref{obsdetails}. We generated exposure maps using \textsc{xisexpmapgen} to identify the regions near the edge of the detectors which are removed due to their low effective area. The calibration regions at the corner of the XIS detectors were removed.

Non X-ray background (NXB) spectra were created using \textsc{xisnxbgen} and were subtracted from the spectra when performing fits. Redistribution matrix files (RMFs) were made using \textsc{xisrmfgen} and Ancilliary Response Files (ARFs) were made using \textsc{xissimarfgen}. We use two ARFs: one assuming a uniform circular source of radius 20$'$ (larger than the detector field of view, FOV) which is used for the uniform background model, and one made using the $\beta$ model best fit to the surface brightness profile found with our \emph{Suzaku} data using the method of \citet{Bautz2009}, which is used for the cluster emission model.   

We remove point sources identified by eye in the \emph{Suzaku} images as shown in Fig. \ref{annuliPKS}, meaning that we resolve the cosmic X-ray background (CXB) as point sources to the same threshold flux (1$\times$ 10$^{-13}$ ergs cm$^{-2}$ s$^{-1}$) across all of the pointings (as is done in \citealt{Bautz2009}, \citetalias{Simionescu2011} and \citealt{Walker2012_A2029}). Point sources below this flux are unresolved and are modelled as a powerlaw component of the background. We derive the expected unresolved CXB flux and its variance for the annular extraction regions later in section \ref{CXB_level}. 

All of the light curves were checked to ensure that no flaring had occurred during the observation in the energy band studied. Following \citet{Humphrey2011} we obtained \emph{ACE} (Advanced Composition Explorer) data using the SWICS (Solar Wind Ion Composition Spectrometer) instrument for the O$^{+7}$/O$^{+6}$ ratio at the times of the observations to gauge the effect of solar wind charge exchange (SWCX) on our observations, following the observation of \citet{Snowden2004} that SWCX is negligible when O$^{+7}$/O$^{+6}$ is less than $\sim$ 0.3. As shown in Fig. \ref{ACE} we find that O$^{+7}$/O$^{+6}$ never increases significantly above $\sim$ 0.2 during the observations, and conclude that SWCX contamination is negligible. As a further check we filter out the time intervals where O$^{+7}$/O$^{+6}$ is greater than 0.2 and find that the spectral fits to the data are not affected. 

All fits were performed in \textsc{xspec} 12.7.1b using the extended C-statistic. We only use data from the front illuminated XIS detectors (XIS0 and XIS3) due to the higher NXB level of the back illuminated XIS1 detector. We fit the XIS0 and XIS3 data for both of the editing modes (3$\times$3 and 5$\times$5) simultaneously (so each fit is done to 4 spectra simultaneously).

\begin{table*}
  \begin{center}
  \caption{Observational parameters of the pointings}
  \label{obsdetails}
  
    \leavevmode
    \begin{tabular}{llllll} \hline \hline
    Name & Obs. ID & Position & Total exposure per detector (ks) & RA & Dec (J2000) \\ 
    & & &(3$\times$3 and 5$\times$5 editing modes) & &\\ \hline
    10 & 802062010 & Centre & 27.9  & 116.8852 & -19.2901 \\
    20 & 802062020 & W & 26.0    & 116.6543 & -19.2063 \\
    30 & 802062030 & N & 26.3    & 116.9737 & -19.0727 \\
    40 & 802062040 & E & 30.6   & 117.1155  & -19.3739 \\
    50 & 802062050 & S & 31.2  & 116.7966 & -19.5079 \\
    53 & 805083010 & Far N & 31.8   & 117.0925 & -18.8008 \\
    54 & 805084010 & NW & 19.3  & 116.7033 & -18.9048 \\
    55 & 805085010 & Far W & 31.1  & 116.3632 &  -19.0871\\
    56 & 805086010 & SW & 23.9  & 116.4685 & -19.4632 \\
    57 & 805087010 & Far S & 32.4  &  116.6522 & -19.7838 \\ \hline
    \end{tabular}
  \end{center}
\end{table*}

\section{Background Subtraction and Modelling}
\label{newbackgroundmodelling}

An accurate understanding of the outskirts of galaxy clusters requires the level, spectral shape and spatial variation of the X-ray background to be understood. The extragalactic component of the X-ray background (the cosmic X-ray background, CXB) which originates from unresolved point sources, and the soft foreground emission from the galaxy both need to be modelled accurately and the magnitude of their expected spatial variations over the cluster properly quantified. 

 We treat each of these background components separately later, but first describe the problems with the background modelling used in G09 and describe the complicated absorbing hydrogen column density in the area around PKS 0745-191 which has not previously been taken into account.

\subsection{Previous Background Subtraction and Modelling}
\label{backgroundmodelling}

In G09 the Lockman Hole was modelled as consisting of a single unabsorbed thermal component (a \textsc{mekal} model) at kT=0.1 keV representing soft galactic emission, added to an absorbed powerlaw of index 1.4 representing the cosmic X-ray background (CXB) from unresolved point sources. This was motivated by the similar count rates of the Lockman Hole and the background region around PKS 0745-191 in the 1-2 keV band as determined from \emph{ROSAT} All Sky Survey (RASS) data. However as PKS 0745-191 lies at a much lower galactic latitude than the Lockman Hole [for PKS 0745-191 (l,b)=(236.4,3.0) and for the Lockman Hole (l,b)=(149.7,53.2)], the absorbing column density is higher in the direction of PKS 0745-191 (with an average of 4.2 $\times$10$^{21}$ cm$^{2}$, compared to 6.2$\times$10$^{19}$ cm$^{2}$ for the Lockman Hole). This means that the actual galactic emission must be higher than that in the direction of the Lockman Hole (to compensate for the higher column density and yield the same count rate), which may be the result of a component not present in the Lockman Hole direction. 

\begin{figure}
  \begin{center}
    \leavevmode
      \epsfig{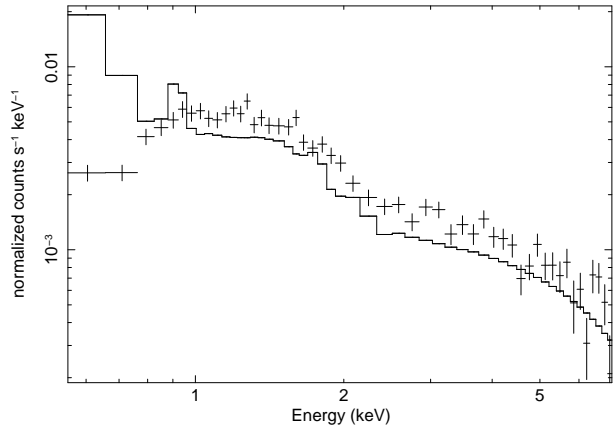}

      \caption{Comparison of the Lockman Hole background model used in G09 with the background spectral data points from the Suzaku background regions (23$'$ to 41$'$) of our PKS 0745-191 pointings. The causes of the discrepancies between the model and the data are described in the main text.} 
      \label{bkgcompare}
  \end{center}
\end{figure}

In Fig. \ref{bkgcompare} we compare the actual background spectrum from the background regions in our new Suzaku observations (regions outside 23$'$) with the Lockman Hole background level used in G09. The CXB level used in G09 (7.26 $\times$ 10$^{-4}$ photons cm$^{-2}$ s$^{-1}$ at 1 keV for a circular area of radius 20 arcmins) corresponds to a 2-10 keV band flux of 1.37$\times$ 10$^{-11}$ ergs cm$^{-2}$ s$^{-1}$, which as we find in section \ref{CXB_level} is at the lower limit of what we would expect given a constant threshold flux of removed point sources of 10$^{-13}$ ergs cm$^{-2}$ s$^{-1}$, causing this background component to be systematically underestimated. This is likely because G09 did not use a constant threshold flux for resolving point sources, and because only one background pointing was used whose area was insufficient to cover a representative sample of point sources in the CXB. As the fit was only performed in the 1-5 keV band, the fit below 1 keV is poor indicating that the background model for the soft emission is incorrect.

\citet{Masui2009} examined a background pointing in the galactic midplane, (l,b)=(235.00,0.00), 3 degrees away from PKS 0745-191, and deduced a significantly different background model to G09 resulting from its lower galactic latitude. In \citet{Masui2009} three background models are presented using two different hypotheses for the background which attempt to explain a narrow bump in the spectrum which peaks at $\sim$ 0.9 keV. The first type, (which is explored in their model A and model B) uses the same CXB and galactic local hot bubble (LHB) model as G09, but includes higher temperature thermal components (for model A one \textsc{vapec} component at 0.766keV is used, while in model B a \textsc{vapec} component at 0.658 keV is added to an \textsc{apec} component at 1.5 keV). In the preferred model (their model C), the hot gas component is replaced with an absorbed bremsstrahlung component at kT$_{C}$=0.183 keV which they use to model the contribution of unresolved faint dM stars to the spectrum (whose number density is higher close to the galactic plane), which is claimed to give a more physical model for the spectrum.  

From Fig. \ref{bkgcompare} we see no significant evidence for the bump in the spectrum at 0.9keV observed in \citet{Masui2009}, and we find that the background can be best fit assuming a model consisting of two thermal components (\textsc{apec}) each of solar metallicity and zero redshift at 0.1 keV (unabsorbed, modelling the LHB) and 0.6 keV (absorbed), and this fit is shown in Fig. \ref{bkg_spectra}. This is most likely because the dM star component which contributes this background component decreases rapidly away from the galactic plane (see Fig 4. of \citealt{Masui2009} which shows the dM star contribution decreasing by a factor of $\sim$ 10 from its galactic midplane value at the latitude of PKS 0745-191, b=3.0), and we find no statistically significant evidence for its presence.

\subsection{Galactic hydrogen column density variations}
\begin{figure*}
  \begin{center}
    \leavevmode
 
        \hbox{
      \includegraphics[width=80mm]{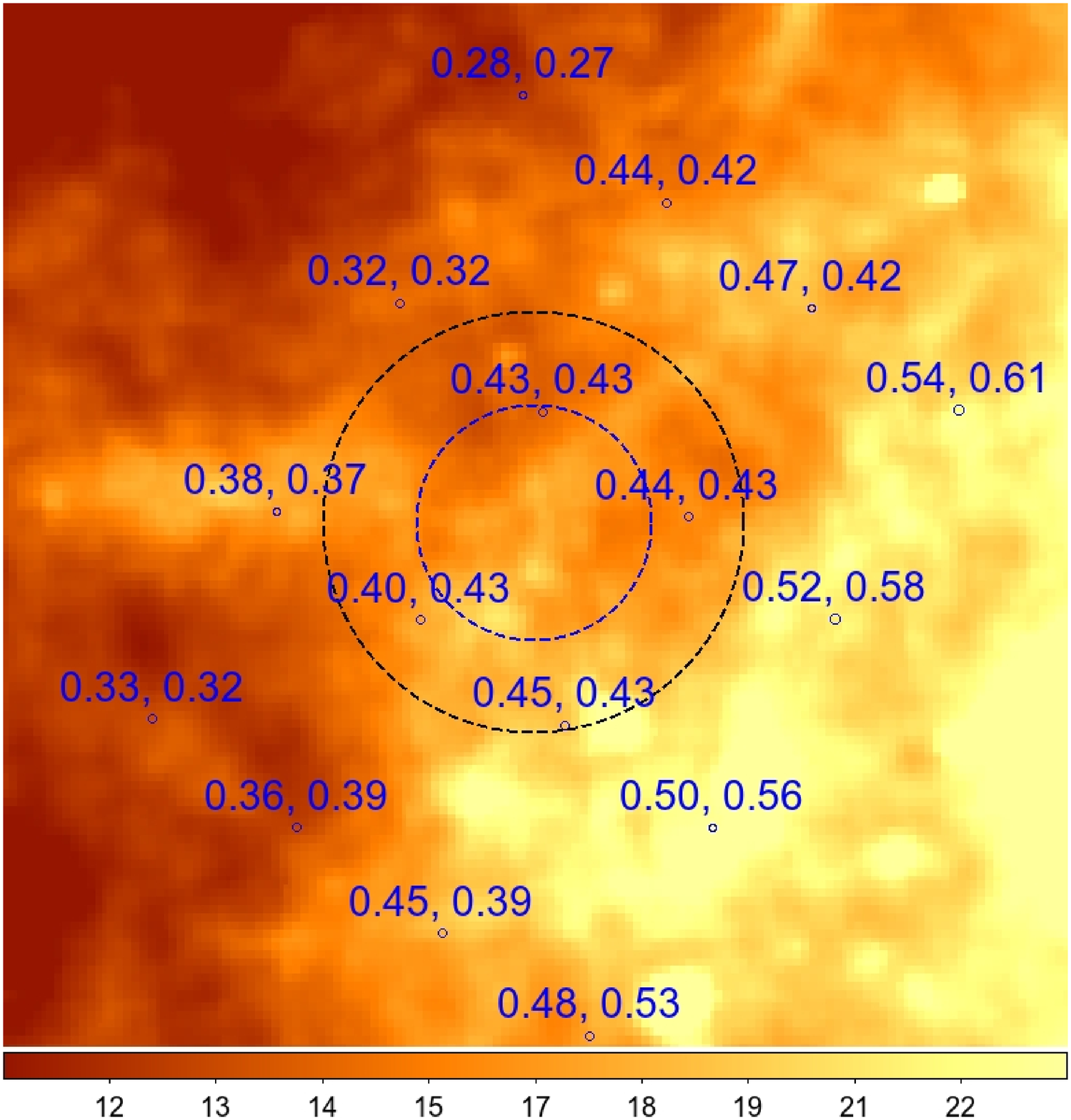}
      \hspace{10mm}
            \includegraphics[width=80mm]{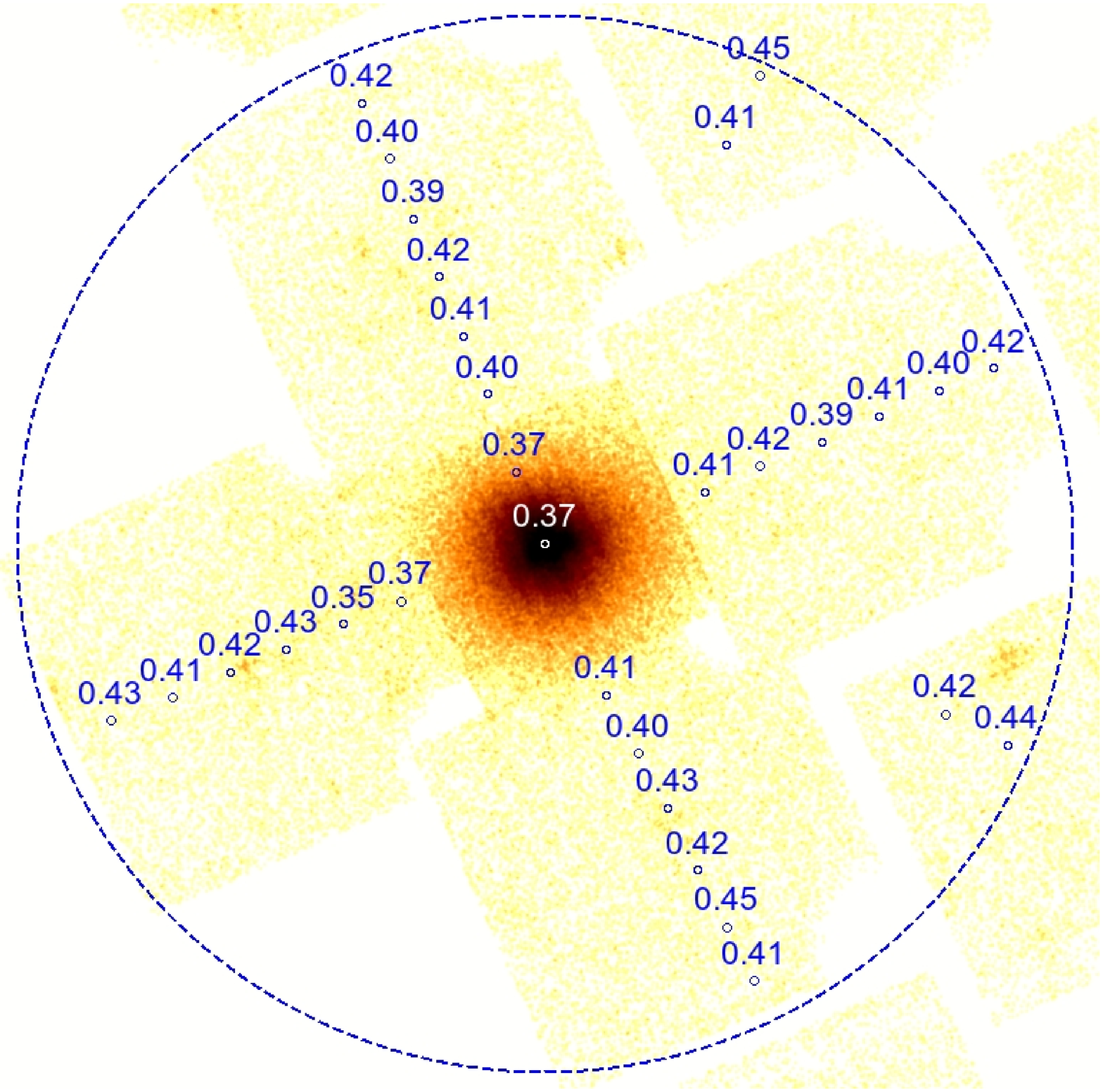}

        }
      \caption{\emph{Left}: IRAS 100 micron map with LAB and Dickey and Lockman hydrogen column density values (units are 10$^{22}$ cm$^{-2}$) overplotted (LAB results are first, then the Dickey and Lockman results). The black dashed circle is at 41$'$ and represents the outer limit of all the pointings. The dashed blue circle of radius 23$'$ is the same in both the left and right panels and represents the boundary beyond which we detect no statistically significant cluster emission. \emph{Right}: Column density values (units are 10$^{22}$ cm$^{-2}$) obtained by letting the column density be a free parameter in our Suzaku fits. The error on each value (combined systematic and statistical error) is $\pm$0.05 $\times$ 10$^{22}$ cm$^{-2}$.} 
      \label{column_density}
  \end{center}
\end{figure*}

The column density of hydrogen in the direction of PKS 0745-191 is high due to its location near to (3 degrees above) the galactic plane, and from the Leiden/Argentine/Bonn (LAB) survey \citep{LABsurvey} and the Dickey and Lockman survey \citep{DL1990} is known to be highly spatially variable, though due to the course sampling of these HI surveys the full extent of the spatial variation is poorly understood. Infrared dust maps such as the 100 micron IRAS dust map shown in Fig. \ref{column_density}, which are known to correlate with column density variations, show a complicated environment not fully represented by the LAB and Dickey and Lockman survey values which are also shown in  Fig. \ref{column_density}. The general trend around PKS 0745-191 shown in Fig. \ref{column_density} is for the column density to increase from east to west, with a range from 0.32 to 0.52 $\times$ 10$^{22}$ cm$^{-2}$ (using the LAB values), and the average value over the cluster is 0.42 $\times$ 10$^{22}$ cm$^{-2}$.

Previous X-ray observations of the bright centre of PKS 0745-191 have found column densities at the lower end of the range indicated by the HI surveys. \citet{Allen2000} studied a \emph{ROSAT} PSPC pointing of PKS 0745-191 and found a best fitting column density at the core of $\sim$ 0.35 $\times$ 10$^{22}$ cm$^{-2}$. \citet{Chen2003} studied \emph{XMM-Newton} pointings and found a best fit central column density of 0.36 $\times$ 10$^{22}$ cm$^{-2}$ with the EPIC-PN detector data and 0.4 $\times$ 10$^{22}$ cm$^{-2}$ with the EPIC-MOS data (this difference is attributed to a systematic calibration problem with these detector's response matrices). As we find later when we allow the column density to be a free parameter in our spectral fits, we find that the column density near the core is consistent with these results from \emph{ROSAT} and \emph{XMM-Newton}, but that away from the core the column density varies spatially (but is still consistent with the range from the HI radio surveys). This is shown in the right panel of Fig. \ref{column_density}.

 \subsection{The CXB}
\label{CXB_level}

The CXB, which consists of unresolved point sources, has been studied extensively and progressively deeper  observations have allowed the point sources from which it is comprised to be resolved to progressively lower flux. \citet{Moretti2003} studied \emph{ROSAT}, \emph{Chandra} and \emph{XMM-Newton} observations and obtained empirical best fits to the cumulative flux distribution (the number of sources have a flux greater than S, N($>$S)) using a two powerlaw model. For studies of the outskirts of galaxy clusters, spatial variations in the CXB due to spatial variations in the number of unresolved point sources presents one factor limiting our ability to detect faint cluster emission in the outskirts. The deeper the CXB can be resolved, the lower the number of unresolved point sources and thus the lower the spatial variations expected in the CXB due to spatial variations in the number of unresolved point sources, making it possible to measure fainter cluster emission with greater statistical significance.

When modelling the expected CXB level and its expected level of fluctuation in the spectral extraction regions, we follow the approach of \citet{Walker2012_A2029}. For the total unresolved CXB flux in the 2-10 keV band, we use the value obtained in \citet{Moretti2009} using \emph{Swift} (2.18 $\pm$0.13 $\times$ 10$^{-11}$ erg cm$^{-2}$ s$^{-1}$ deg$^{-2}$), which is in complete agreement with the value determined in \citet{Moretti2003} (2.24 $\pm$ 0.16 $\times$ 10$^{-11}$ erg cm$^{-2}$ s$^{-1}$ deg$^{-2}$). We avoid using \emph{ASCA} results of \citet{Kushino2002} due to the stray light problems from which \emph{ASCA} suffered. 

To determine the amount of the CXB resolved, we integrate the cumulative flux distribution above the threshold flux to which our \emph{Suzaku} observations allow point sources to be resolved (S$_{excl}$ = 10$^{-13}$ erg cm$^{-2}$ s$^{-1}$ in the 2-10 keV band) and subtract this from the total 2-10 keV band flux (in ergs cm$^{-2}$ s$^{-1}$ deg$^{-2}$) as shown below, 

\begin{eqnarray}
F_{\rm CXB} = 2.18 \pm 0.13 \times 10^{-11} - \int_{S_{\rm excl}}^{S_{\rm max}}
\Big(\frac{dN}{dS} \Big)
\times S ~ dS
\end{eqnarray}

which leads to an unresolved 2-10 keV band CXB flux of 1.87$\pm$0.13 $\times$ 10$^{-11}$ erg
cm$^{-2}$ s$^{-1}$ deg$^{-2}$.

Following \citet{Bautz2009}, we then calculate the expected spatial variations in the 2-10 keV band CXB flux due to the unresolved point sources in each annular spectral extraction region (of solid angle $\Omega$) using 

\begin{eqnarray}
\sigma^2_{\rm CXB} = (1/\Omega) \int_{0}^{S_{\rm excl}} \Big(\frac{dN}{dS}
\Big)
\times S^2 ~ dS
\end{eqnarray}

from which we can see that the variance can be minimised by using larger extraction regions and a smaller threshold flux (S$_{\rm excl}$).

The resulting one sigma uncertainties for each annulus in which we investigate cluster emission are tabulated in table \ref{CXB_fluctuations}, and these are the ranges by which we must vary the CXB background model to understand the systematic error on the ICM measurements of our uncertainty on the CXB level. These predicted variations are entirely consistent with those derived by scaling the fluctuations observed by Ginga \citep{Hoshino2010}, and in the \emph{Suzaku} background observations of the Perseus cluster \citepalias{Simionescu2011conf}.

\begin{table*}
  \begin{center}
  \caption{Expected CXB one sigma variations in each annulus in the 2-10 keV band (10$^{-12}$ erg s$^{-1}$
cm$^{-2}$ deg$^{-2}$)}
  \label{CXB_fluctuations}

    \leavevmode
    \begin{tabular}{llllllll} \hline \hline

           0.0$'$-2.5$'$ & 2.5$'$-6.0$'$ & 6.0$'$-8.8$'$ & 8.8$'$-11.7$'$ & 11.7$'$-14.5$'$ & 14.5$'$-17.3$'$ & 17.3$'$-20.2$'$ & 20.2$'$-23$'$    \\ \hline
                 6.7    &  3.1    &  3.1     & 2.5   &   2.5   &   2.4   &   2.1 &     2.3 \\    \hline
    \end{tabular}
  \end{center}
\end{table*}

\subsection{Modelling soft foreground.}
\label{softforeground}

The soft foreground resulting from galactic emission is challenging to measure for PKS 0745-191 due to its location near the galactic plane and the highly spatially
variable hydrogen column density indicated by the HI survey maps and the IRAS 100 micron dust maps. Therefore, in addition to spatial variation of the galactic emission, there is
also spatial variation of the column density absorbing this emission. 

When fitting for the soft foreground using the Suzaku observations outside 23$'$, we find it is best modelled using two \textsc{apec} components each with solar metallicity and zero redshift (one at 0.6 keV which is absorbed by the galactic column, and one at 0.1 keV which is unabsorbed and corresponds to emission from the Local Hot Bubble). The fit is not improved by including another low temperature absorbed component corresponding to the hot halo surrounding the galaxy which is often used \citepalias{Simionescu2011conf}. In general it is not possible to fit for the \textsc{apec} normalisations of the galactic background components and the column density simultaneously owing to the strong degeneracy between them, in that
the same background emission level is achieved by reducing both the column density (i.e. less absorption) and the galactic \textsc{apec} normalisations (i.e. less galactic emission).

To begin with we fit the 32$'$-41$'$ regions of the Suzaku pointings with the background model described earlier, initially fixing the column density to the mean value over the cluster of 0.42$\times$ 10$^{22}$ cm$^{-2}$ and allowing the \textsc{apec} and CXB normalisations to be free parameters. We then compare the 23$'$-32$'$ regions in the new pointings with them to see whether we can see evidence for cluster emission. We find fluctuations in the soft band flux, but these can be explained by variations in the column density which are in the range expected from the LAB survey values. So there is no statistically significant evidence for excess emission above the background in the region outside 23$'$.

 Thus we find that although the soft background varies between 23$'$ and 41$'$, it is consistent with the simple possibility of uniform galactic emission absorbed by a varying column density within the range
of the LAB survey values. However the possibility remains that the intrinsic galactic emission and the column density are both varying, so for example it is not possible to say
with certainty that a background region with higher than average soft band flux is a region with lower than average galactic column; it could instead be a region 
with higher than average galactic emission with the same absorbing column as other regions. It is not possible to break this degeneracy due to the courseness
and uncertainties of the LAB survey galactic hydrogen column values. 

Due to this, we adopt a conservative approach when modelling the soft foreground emission which fully encompasses our uncertainty in the spatial variation of the galactic emission and the spatial variation of the column density absorbing it. The Suzaku background regions are divided into annuli between 23$'$-28$'$, 28$'$-32$'$ and 32$'$-41$'$ (yielding 13 regions in total as shown in Fig. \ref{annuliPKS}) so that the extraction regions are approximately the same size as the cluster emission annuli, allowing us to understand the expected variation of the soft foreground on this distance scale. We fit each Suzaku
background region with the phabs(apec(0.6keV)+powerlaw)+apec(0.1keV) model using the minimum absorbing column indicated by the LAB survey (0.32$\times$ 10$^{22}$ cm$^{-2}$) and then for the highest absorbing column
(0.52$\times$ 10$^{22}$ cm$^{-2}$). The galactic \textsc{apec} and powerlaw norms are free parameters which we fit for. From this we obtain the maximum and minimum galactic \textsc{apec} norms for each component (the highest coming for the higher absorbing column), and this gives the full range within 
which we can say with certainty the intrinsic galactic \textsc{apec} norms must lie, and this range folds together our uncertainty in the spatial variation of the absorbing column and uncertainty 
in the intrinsic galactic emission. For all of the fits the CXB level is consistent within the expected variance with that calculated in section \ref{CXB_level} using the results of \citet{Moretti2003} and the point source threshold flux of 10$^{-13}$ erg cm$^{-2}$ s$^{-1}$ in the 2-10 keV band.

As a further check we perform spectral fits to the 25$'$-50$'$ annulus of the ROSAT PSPC pointing for PKS 0745-191 (obs ID rp800623n00, which is shown in Fig. \ref{ROSATimage}), dividing it into 4 sectors to the north, south, east and west, allowing us to measure the soft background in the east which is not covered by our \emph{Suzaku} background pointings. The ROSAT data were reduced and filtered using the standard procedures described in the document 'ROSAT data analysis using xselect and ftools' \footnote{heasarc.gsfc.nasa.gov/docs/rosat/ros\_xselect\_guide/xselect\_ftools.html}. Particle background spectra were modelled using the ftool \textsc{pcparpha} and subtracted, and ROSAT ARFs were created using the ftool \textsc{pcarf}. The galactic \textsc{apec} components are found to be in agreement with the values from the \emph{Suzaku} background pointings, and the background level of the east direction is consistent with that of the other directions. 

We take the best fitting galactic \textsc{apec} norms to be the values obtained when fitting all of the 13 background regions simultaneously with a column density of 0.42$\times$10$^{22}$ cm$^{-2}$ (the mean column density from the LAB survey) and these are norm(0.1 keV)=0.015 cm$^{-5}$ and norm(0.6 keV)=0.0013 cm$^{-5}$, where the normalisations have units of $10^{-14}(4\pi)^{-1}D_{A}^{-2}(1 + z)^{-2} \int n_{e} n_{H} dV$, where   $D_{A}$ is the angular size distance (cm), and $n_{e}$ and $n_{H}$ are the electron and hydrogen densities (cm$^{-3}$) respectively, and these values are scaled for a circular area of sky of 20$'$ radius (1257 arcmin$^{2}$). The range for each norm found by looking at all 13 Suzaku background regions and using a column density range between 0.32-0.52$\times$10$^{22}$ cm$^{-2}$ was 0.01-0.02 cm$^{-5}$ for the 0.1keV norm and 0.0009-0.0017 cm$^{-5}$ for the 0.6 keV norm. These ranges include the effect of accounting for stray light emission in the background regions which is modelled using the ray tracing simulator \textsc{xissim} and subtracted. 

When performing the initial spectral fits to the cluster emission we use the best fitting galactic \textsc{apec} norm values for the galactic background obtained using the average column density of 0.42$\times$10$^{22}$ cm$^{-2}$, and then later investigate the systematic error on the fits 
of varying all of the background components (including the CXB and the NXB) by their quantified uncertainties while letting the column density be a free parameter. This is described in detail in section \ref{systematicerrors}.

\subsection{Stray Light and PSF spreading}
\label{stray light}
\emph{Suzaku} observations suffer from stray light effects due to X-rays passing through the telescopes along undesired optical paths, such as reflection purely from the secondary mirror and reflection from the back surface of the primary mirror \citep{Serlemitsos2007}. This results in X-rays being detected from bright sources outside of the field of view of the telescope. In addition there is also light spread from adjacent regions due to Suzaku's broad PSF.

We estimate the relative contributions into each annulus due to the PSF and stray light as follows. Using the best fit Suzaku surface brightness profile found later in section \ref{sbcomparison} for each annulus as input to the ray tracing simulator \textsc{xissim} using the same number of photons (2$\times$10$^{6}$), we measure the number of photons scattered into the other annuli, and calculate the fractional contamination in each annulus from all of the other annuli. This is shown in table \ref{psf_table}. An error in the calculation of the mixing fractions in G09 led to the
contributions entering the outer annuli from the inner annuli being
underestimated in table 2 of G09, (for instance the
contribution from the core into the outermost annulus in G09 is incorrectly
estimated to be 0.1 percent). The fractional contributions 
of stray light entering the outer annuli we calculate are similar to levels that have been measured in other \emph{Suzaku} studies \citep{Bautz2009, Hoshino2010,  Akamatsu2011, Walker2012_A2029}.

\begin{table*}
  \begin{center}
  \caption{Percentage contribution of flux in the rows' annulus from the columns' annulus due to PSF spreading and stray light.}
  \label{psf_table}
  
    \leavevmode
    \begin{tabular}{lllllllll} \hline \hline
     
    & 0.0$'$-2.5$'$  & 	2.5$'$-6.0$'$	 & 6.0$'$-8.8$'$ & 	8.8$'$-11.2$'$	 & 11.2$'$-14.5$'$  & 14.5$'$-17.3$'$  & 17.2$'$-20.1$'$ &  20.1$'$-23.0$'$\\ \hline
0.0$'$-2.5$'$    &    96.6    &    3.4   &   0.01   &  0.002  &  0.0003  &  0.0002  &  0.0002  &  5$\times$10$^{-5}$\\
2.5$'$-6.0$'$     &   38.5    &    60.5    &   1.0   &   0.03   &  0.008    & 0.003  &   0.002   &  0.001\\
 6.0$'$-8.8$'$    &   16.8    &    14.1    &    64.6   &     4.0   &    0.43    &   0.1   &   0.04    &  0.02\\
8.8$'$-11.2$'$     &   12.7    &    7.0    &    8.9    &    64.9   &     5.8    &   0.6   &    0.1   &   0.03\\
11.2$'$-14.5$'$    &    5.5   &     4.4   &     1.8   &     12.9   &     64.7   &     10.1  &     0.5   &    0.1\\
14.5$'$-17.3$'$    &    1.0   &     1.4  &     1.0   &     1.7   &     14.7   &     72.1   &    7.6   &    0.4\\
17.2$'$-20.1$'$    &    2.9   &     1.1   &    0.6   &     1.2   &     2.1   &     19.9     &   67.0    &    5.2\\
20.1$'$-23.0$'$    &    8.7    &    1.9   &    0.4   &    0.7   &     1.4    &    2.5     &   16.1    &    68.4\\ \hline

    \end{tabular}
  \end{center}
\end{table*}

We see that the majority of the emission observed in each annulus originates from that annulus, and the majority of the contamination originates from the neighbouring annuli which will have similar temperatures and so the effect of this should be low. Contamination from the bright central 6$'$ is non-negligible, and so when performing fits to the regions outside 6.0$'$ we investigate the systematic error of including in the background model a model of the emission reaching those regions from the central 6$'$ (see section \ref{systematicerrors}). This was achieved by using \textsc{xissim} to simulate the emission in all of the pointings using the spectrum, image and flux of the central 6$'$ as input.

\section{Spectral Analysis}
\label{analysis}

To investigate the cluster emission we divide the pointings into the annuli shown in Fig. \ref{annuliPKS}, with the annuli between 0.0$'$-2.5$'$, 2.5$'$-6.0$'$, 6.0$'$-8.8$'$, 8.8$'$-11.7$'$, 11.7$'$-14.5$'$, 14.5$'$-17.3$'$, 17.3$'$-20.2$'$, 20.2$'$-23$'$, all of which are at least as wide as the Suzaku PSF (2.5$'$) to ensure that the majority of the flux measured in each annulus originated from that annulus. We have already concluded earlier that there is no statistically significant emission above the background level for radii greater than 23$'$.

\subsection{Azimuthally averaged deprojected fits}
We initially fit each direction in the cross separately but found no statistically significant azimuthal variation in the temperature profiles, so we focus on azimuthally averaged profiles. We find the azimuthally averaged deprojected temperature and density profiles over the available pointings. We perform fits in the 0.7-7.0 keV band and allow the metallicity to be free for the central 3 annuli, but fix it to 0.3 Z$_{\odot}$ for the outer 5 annuli which is commonly found and was used in G09 and E11a. For the central 3 annuli we find metallicities of 0.37$^{+0.05}_{-0.05}$ Z$_{\odot}$, 0.34$^{+0.08}_{-0.08}$ Z$_{\odot}$ and 0.2$^{+0.1}_{-0.1}$ Z$_{\odot}$, consistent with the values found in \citet{Snowden2008} using \emph{XMM-Newton}. 

We allow the column density to be a free parameter for each region to accommodate for its spatial variation. The values for the column density found by letting this parameter be free are shown in Fig. \ref{column_density}, and are all within the range indicated by the LAB survey. The spectral fits are shown in Fig. \ref{cluster_spectra}. 

\begin{figure}
  \begin{center}
    \leavevmode
    \hbox{
      \epsfig{figure=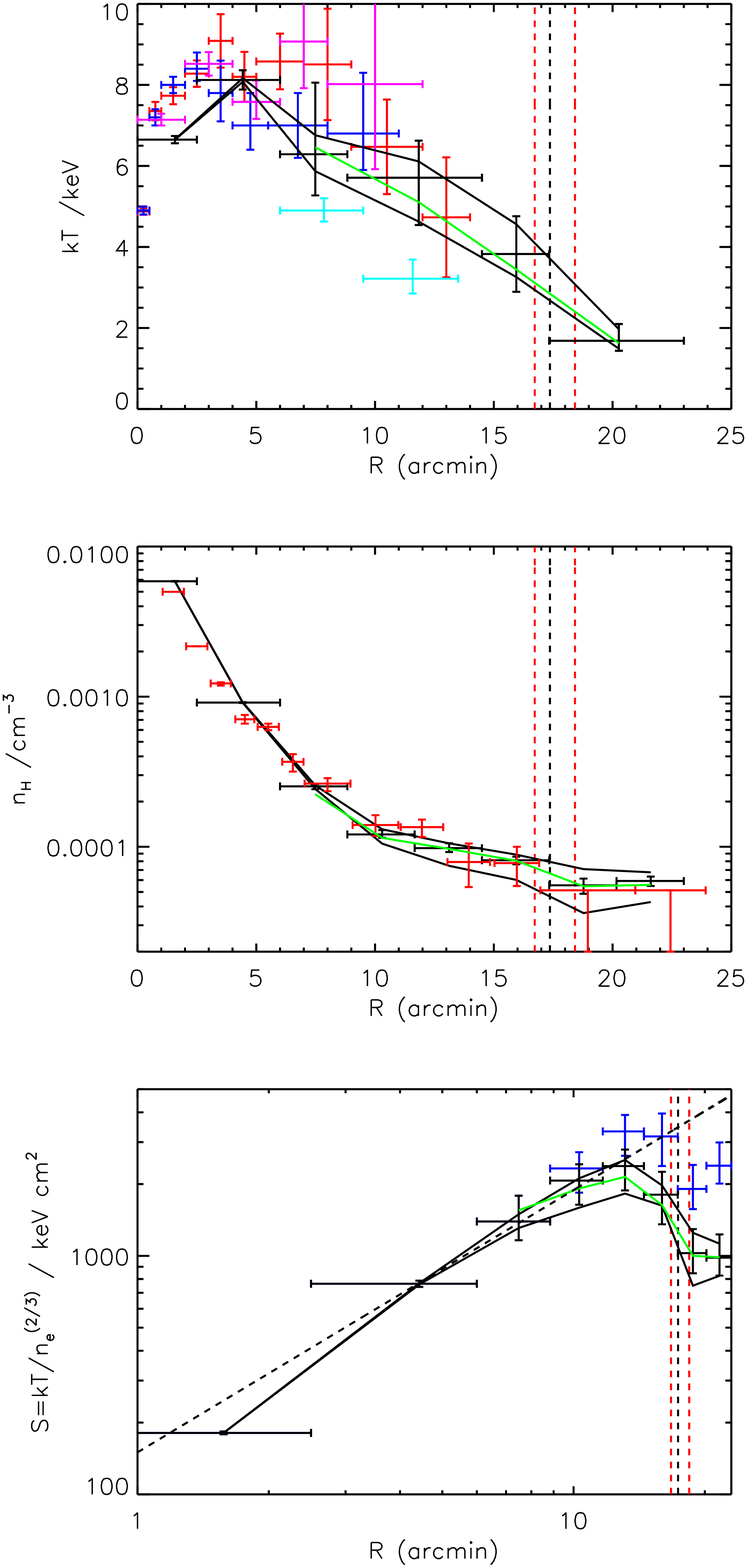, angle=0,
        width=0.95\linewidth}
        }

      \caption{\emph{Top}: Deprojected, azimuthally averaged temperature profile from Suzaku shown with black points. Blue points are from \emph{Swift} \citep{Moretti2011}, the red points are from \emph{XMM-Newton} \citep{Snowden2008} and the pink points from from \emph{BeppoSAX} \citep{DeGrandi2002}, showing good agreement with previous observations. The projected temperatures obtained in G09 in the 6.0$'$-13.5$'$ region are shown as the cyan points, and are significantly lower as described in E11a. \emph{Middle}: Deprojected density profile from Suzaku is in black. The deprojected \emph{ROSAT} density profile from E11a is in red, where the outer 2 density values are 90 percent upper limits.  \emph{Bottom}: Entropy profile obtained using deprojected densities and deprojected temperatures. The entropies corrected by the clumping factors found later in Fig. \ref{gasmassprofile} are shown as the blue points. The dashed black line shows the theoretical prediction (S $\propto$ r$^{1.1}$) from the numerical simulations of \citet{Voit2005}. One sigma systematic errors owing to background modelling are shown as the solid black lines for the temperature, density and entropy profiles. The solid green line show the best fits obtained when the stray light from the central 6$'$ is subtracted from the cluster spectra.  }
      \label{T_and_d_profiles}
  \end{center}
\end{figure}

The deprojected temperature, density and entropy (S=kT/n$_{e}^{2/3}$) profiles are shown in Fig. \ref{T_and_d_profiles}. We model each annulus as the sum of the absorbed \textsc{apec} components describing emission originating from the annulus itself and from the shells outside it, scaling each \textsc{apec} component according to the volumes projected onto the annulus from the external shells assuming spherical symmetry. This emulates the \textsc{xspec} mixing model \textsc{projct}, and allows data from different detectors to be fitted simultaneously (this is the same method described in \citealt{Humphrey2011} and \citealt{Walker2012_A2029}). We rebin the temperature profile with more course binning owing to the limited spectral quality of the data, which is achieved by tying together the temperatures of the two annuli between 8.8$'$ and 14.5$'$, and the two annuli between 17.3$'$ and 23.0$'$. We fit all 8 annuli simultaneously to propagate errors.

\subsection{Systematic errors}
\label{systematicerrors}
We now turn to investigate systematic errors in the measurements. We need to quantify the systematic error on the temperature and density profiles resulting from our uncertainty in the background model used. As we found in section \ref{backgroundmodelling}, the uncertainties in the variation of the column density increases the uncertainty in the level of the galactic emission, and we found that the normalisations of the 0.1 keV and 0.6 keV \textsc{apec} components could only be confined within the ranges of norm(0.1keV)=0.010-0.020 cm$^{-5}$ and norm(0.6keV)=0.0009-0.0017 cm$^{-5}$. The CXB is also expected to vary spatially owing to variations in the number of unresolved point sources in each annulus, and the one sigma variations from the expected 2-10 keV band flux of 1.87$\pm$0.13$\times$10$^{-11}$ erg
cm$^{-2}$ s$^{-1}$ deg$^{-2}$ for each annulus have already been calculated in table \ref{CXB_fluctuations}. In addition to this the NXB level has an uncertainty of $\pm$3 percent \citet{Tawa2008}.

Previous studies of cluster outskirts (for instance G09 and \citealt{Akamatsu2011}) have quantified the effect of these background uncertainties by varying each parameter by its one sigma error and measuring the effect on the temperature and densities. This approach is overly simplistic and underestimates the true systematic error caused by the uncertainty on the background model. A more realistic approach requires that all of the background components are varied simultaneously through their expected variances, and the effect on the temperature and density profiles measured. 

To achieve this we follow the method of \citet{Walker2012_A2029} and produce 10000 realisations of the background model (the galactic \textsc{apec} norms, the CXB norms and the NXB level) distributed by their expected variations which we have already found, and accounting for covariance between the background parameters. We then perform the deprojected fit for each realisation of the background, thus folding our uncertainty in the background modelling through the deprojection to completely propagate the errors. From the resulting 10000 temperature and density profiles we obtain, the one sigma systematic errors are found for each annulus by finding the ranges which encompass 68 percent of the values around the peak value. These systematic errors are plotted as the solid black lines in Fig. \ref{T_and_d_profiles}. 

We then perform the deprojected fits including the stray light emission from the central 6$'$ (the bright core) as a background component, and plot the effect of this on the temperature and density profiles as the green line in Fig. \ref{T_and_d_profiles}. We find that taking into account the stray light is a small effect which is well within the statistical errors of the fits. We vary the contamination on the optical blocking filter (OBF) by $\pm$10 percent (as in \citealt{Akamatsu2011}), using the ftool \textsc{xiscontamicalc} to modify the ARFs, and find that this has a negligible effect which is much smaller than the statistical errors on the temperatures and densities. 

\subsection{Discussion of temperature, density and entropy profiles}
\label{discussprofiles}
As mentioned in E11a, the temperature profile found in G09 is significantly lower than that obtained 
using \emph{XMM-Newton} \citep{Snowden2008} and \emph{Swift} \citep{Moretti2011} outside 6 arcmins, leading to a lower mass and r$_{200}$ measurement than has been obtained
with \emph{XMM-Newton} and \emph{Swift}. We find that our temperature profile in the range 6.0$'$-14.5$'$ is significantly higher than that obtained in G09 and is consistent 
with the \emph{Swift} and \emph{XMM-Newton} results. This is shown in the top panel of Fig. \ref{T_and_d_profiles}. This
is a result of a combination of more accurate background modelling than that used in G09, and also of letting the column density be a free parameter. G09 fixed the
column density to 0.37$\times$10$^{22}$ cm$^{-2}$, which was found to be the best fitting value when the column density was allowed to be free for the the central region (within 2.5$'$). 

For comparison, the \emph{Swift} results of \citet{Moretti2011} fix the column density to the mean LAB survey value of 0.418$\times$10$^{22}$ cm$^{-2}$ for the entire cluster. The 
column density used in \citet{Snowden2008} for the \emph{XMM-Newton} results was allowed to be a free parameter for each annulus studied. 

The deprojected density profile agrees well with that obtained with \emph{ROSAT} in E11a out to 17$'$, and is consistent with the 90 percent upper limit beyond this when the systematic errors in the background modelling are taken into account. As described in detail later in section \ref{sbcomparison}, the ROSAT surface brightness (and therefore density) in E11a appears to have been underestimated in the outskirts due to inconsistent point source removal which has caused the background to be overestimated in E11a. As no emission is found outside 23$'$ the excess cannot be the result of projected emission from outside the 20.1$'$-23$'$ annulus. Including the emission spread into the annulus from neighbouring annuli due to the PSF has no statistically significant effect.

We find the entropy (S=kT/n$_{e}^{2/3}$) profile using the deprojected temperatures and deprojected densities as shown in the bottom panel of  Fig. \ref{T_and_d_profiles}, and compare it to the theoretical prediction of S $\propto$ r$^{1.1}$ obtained in \citet{Voit2005} assuming purely gravitational hierarchical structure formation (and neglecting non-gravitational processes such as cooling and feedback). The higher temperature found in the 6.0$'$-14.5$'$ region compared to G09 causes the entropy profile to be consistent with the S $\propto$ r$^{1.1}$ relation out to greater radius than in G09 (15$'$ compared to $\sim$10$'$).  Outside 15$'$ the entropy profile flattens and is much lower than the S $\propto$ r$^{1.1}$ relation, which is due to the low outer temperature ($\sim$ 2keV). This entropy profile flattening has been observed in other clusters observed to the virial radius with Suzaku, namely Abell 1689 \citep{Kawaharada2010}, Abell 1795 \citep{Bautz2009}, the Perseus cluster \citepalias{Simionescu2011} and Abell 2142 \citep{Akamatsu2011}), and may indicate that the ICM is out of hydrostatic equilibrium due to the continual infall of matter onto the cluster (which has not been virialised yet) as it continues to form. In section \ref{massanalysis} we find that r$_{200}$ (the boundary within which we expect the cluster to be virialised) lies between 16.7$'$ and 18.4$'$, which is consistent with the suggestion that the cold, low entropy gas we find outside 17$'$ is out of hydrostatic equilibrium and has not yet been virialised as it accretes onto the cluster.  

\begin{figure*}
  \begin{center}
    \leavevmode

      \includegraphics[width=100mm]{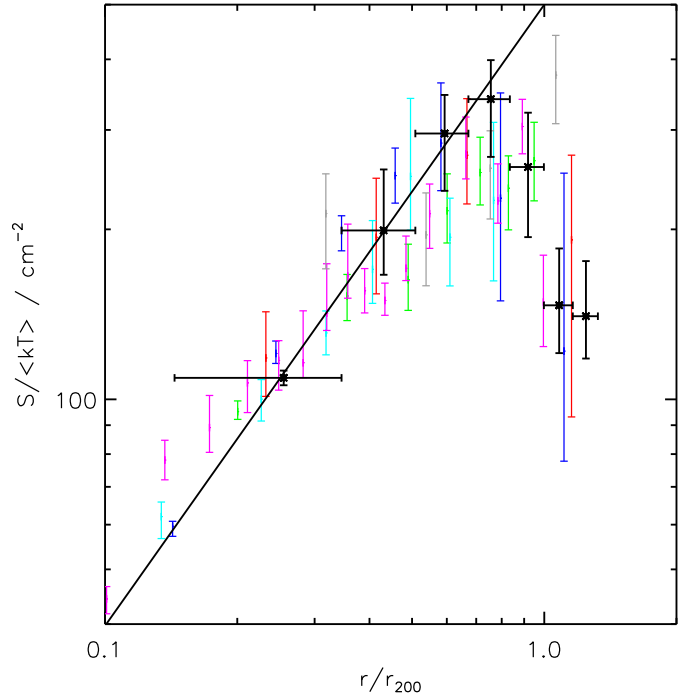}
        
      \caption{Azimuthally averaged entropy profiles scaled by the average ICM temperature for clusters investigated by Suzaku to r$_{200}$, showing a low level of scatter. The black crosses are our new results for PKS 0745-191. The radial error bars for the other clusters are omitted to improve figure clarity. Green is for Abell 2029 \citep{Walker2012_A2029}, blue is for Abell 2142 \citep{Akamatsu2011}, red is for Abell 1689 \citep{Kawaharada2010}, cyan is for Hydra A \citep{Sato2012}, grey is for Abell 1413 \citep{Hoshino2010} and pink is for Perseus (NW direction only from \citetalias{Simionescu2011}). The solid black line shows the S $\propto$ r$^{1.1}$ relation from \citet{Voit2005}.} 
      \label{entropy_compare}
  \end{center}
\end{figure*}

In Fig. \ref{entropy_compare}, we compare the entropy profile (scaled by the average ICM temperature) with the entropy profiles found for other clusters investigated by Suzaku to r$_{200}$. \citet{Sato2012} has shown that when scaled in this way the entropy profiles are consistent with a low level of scatter, but that the PKS 0745-191 results from G09 are lower and inconsistent with the trend. Our new entropy profile shows much stronger agreement with the general trend, resulting mostly from the higher temperatures we find between 6.0$'$-14.5$'$.

\section{Mass Analysis}
\label{massanalysis}

We assume that the ICM is spherically symmetric and is in hydrostatic equilibrium, so that it obeys \citep{Vikhlinin2006}, 

\begin{equation}
\label{hydroeq}
M(<r)=-\mathrm{3.68}\times\mathrm{10}^{\mathrm{13}}M_{\odot}T(r)r\left(\frac{d \rm \:ln \:n_{\rm
    H}}{d \:\mathrm{ln \:r}} + \frac{d\rm\:ln\:T}{d\rm\:ln\:r}\right) \\
\end{equation}

Furthermore, we assume that the total mass density obeys a Navarro Frenk White (NFW) profile \citep{NFW1997} (this assumption is strongly motivated by numerical simulations of hierarchical structure formation)
\begin{eqnarray}
\rho(r)&=&\frac{\rho_0}{r/r_{\rm s}\left(1+(r/r_{\rm s})\right)^2} \\
\rho_0 &=& 200\rho_c c_{200}^{3}/3(ln(1+c_{200}) - c_{200}/(1+ c_{200})) 
\end{eqnarray}

where $\rho_{c}$ is the critical density of the universe, and c$_{200}$=r$_{200}$/r$_{s}$. An in G09, we follow a mass analysis technique similar to that used in \citet{Schmidt2007}. Using the deprojected density profile and the NFW mass profile we calculate the 
temperatures in each annulus, beginning with the outermost annulus, and compare them to the observed temperatures. The results are unchanged if we start with the innermost annulus. The best fit NFW mass profile corresponds to that which produces temperatures which 
match the observed temperatures most closely, so that it minimises the $\chi^{2}$ statistic defined as
\begin{eqnarray}
\label{chisqequation}
\chi^{2} = \sum_{i}\dfrac{(T_{calculated,i} -T_{actual,i})^2}{\sigma_{T_{actual,i}}^{2}}
\end{eqnarray}

\begin{figure*}
  \begin{center}
    \leavevmode
\hbox{
      \epsfig{figure=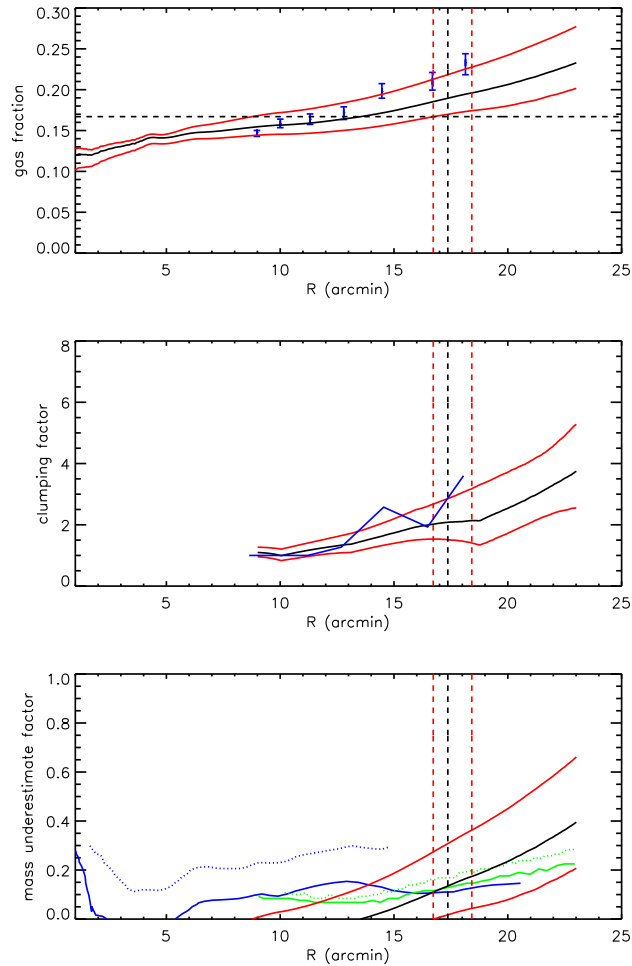, angle=0,
        width=0.5\linewidth}
         }
        
      \caption{  \emph{Top panel}: Gas mass fraction profile; the horizontal dashed black line shows the cosmic mean baryon fraction determined in \citet{Komatsu2011}. The blue points are those obtained for the Perseus cluster in \citetalias{Simionescu2011} scaled to the r$_{200}$ value we find for PKS 0745-191. \emph{Middle panel}: Clumping factor by which the gas density must be overestimated if the true cumulative gas mass fraction does not increase above the mean cosmic baryon fraction. The blue line is the clumping factor determined in \citetalias{Simionescu2011} for the Perseus cluster.   \emph{Bottom panel}: The black line shows the factor by which the total mass needs to be underestimated for the cumulative gas mass fraction to remain below the mean cosmic baryon fraction, with the red lines showing the one sigma error. The green and blue sold lines show the expected mass underestimates from the the numerical simulations of \citet{Lau2009} and \citet{Burns2010} respectively, while the dashed green and blue lines show the upper limit from these simulations.  \emph{All panels}: The best fit is in black and the red lines show one sigma errors. r$_{200}$ is shown by the vertical dashed black line and its one sigma errors by the vertical dashed red lines.}
      \label{gasmassprofile}
  \end{center}
\end{figure*}

 We generate 10000 realisations of the density profile distributed by the combined statistical and systematic errors shown in Fig. \ref{T_and_d_profiles}
and repeat the mass analysis for each profile, calculating the cumulative gas mass fraction profile for each iteration. The resulting gas mass fraction profile is shown in Fig. \ref{gasmassprofile} along with its 1 $\sigma$ errors. To resolve the central annulus, which is needed for constraining the NFW profile, we use the \emph{Swift} temperatures and densities for 
the regions within 2$'$ obtained in \citet{Moretti2011}.

 We find that our higher Suzaku temperatures and more accurate error measurements compared to G09 allow a much better fit to an NFW profile to be obtained, with a best fit
$\chi^{2}$/d.o.f of 9.0/7 (compared to 39/7 obtained in G09) when we used the entire temperature profile except for the outermost temperature annulus. We do not include the outermost temperature annulus in the mass analysis because when we perform a simple mass analysis using linear gradients between the temperature points in the hydrostatic equilibrium equation we find that the hydrostatic mass unphysically decreases in the outermost annulus due to the low temperature, indicating that the ICM is out of hydrostatic equilibrium there. Including this low temperature outer annulus gives a much poorer fit to an NFW profile ($\chi^{2}$/d.o.f = 14.0/8), which, together with the flattening of the entropy profile there, adds further evidence that the ICM is out of hydrostatic equilibrium there. This is similar to the case of Abell 1689 studied in \citet{Kawaharada2010}, where the directions in which the entropy profile flattened were found to have mass profiles which unphysically decreased in the outskirts, indicating the breakdown of hydrostatic equilibrium. 

When the 14.5$'$-17.3$'$ temperature annulus was removed from the mass analysis the NFW
parameters were not significantly changed, resulting in the same r$_{200}$ and M$_{200}$
values. This suggests that in the
14.5$'$-17.3$'$ annulus the ICM is still in hydrostatic equilibrium, and that it is
only outside $\sim$17.3 arcmins that deviations from hydrostatic equilibrium or
non-thermal pressure support become significant.

We find M$_{200}$=9.6$^{+1.9}_{-1.0}$ $\times$10$^{14}$ M$_{\odot}$, c$_{200}$=5.3$^{+0.6}_{-0.9}$, r$_{s}$=370$^{+120}_{-50}$ kpc, 
and r$_{200}$=2.0$^{+0.1}_{-0.1}$ Mpc (17.4$^{+1.0}_{-0.7}$ arcmins), which are consistent with the results obtained for PKS 0745-191 in \citet{Pointecouteau2005}, \citet{Voigt2006} and 
\citet{Schmidt2007}, and with the scaling relations of \citet{Arnaud2005}, all of which are tabulated in table \ref{compare_masses}. It must be stressed however that the \emph{Chandra} results of
\citet{Voigt2006} and \citet{Schmidt2007} used temperature profiles extending
out to only $\sim$400 kpc ($\sim$3.5 arcmins), while the
\emph{XMM-Newton} results
of \citet{Pointecouteau2005} used a temperature profile which only extended to
$\sim$1000kpc ($\sim$9 arcmins), and that these were extrapolated outwards to find M$_{200}$.

We see therefore that the lower temperatures obtained in G09 due to incorrect background modelling caused the mass determination in G09 to be too low and the NFW concentration parameter to be 
unusually higher than previous measurements. 

We find that the gas mass fraction rises with increasing radius and rises above the mean cosmic baryon fraction obtained using CMB data in \citet{Komatsu2011} from around r$_{200}$ onwards, but is lower than (but still consistent with) the cumulative gas mass fraction profile for the Perseus cluster in \citetalias{Simionescu2011} (scaled to the r$_{200}$ value we find), as shown in Fig. \ref{gasmassprofile} . If we assume that the cumulative gas mass fraction is overestimated due to gas clumping alone then the clumping factor (by which the gas density is overestimated) which is required for the cumulative gas mass fraction not to exceed the cosmic mean baryon fraction is shown in the middle panel of Fig. \ref{gasmassprofile} and is lower than that found in \citetalias{Simionescu2011} for the Perseus cluster, but is still well within range indicated by the simulations of \citet{Nagai2011}.  

Assuming that the gas density has been overestimated by the clumping factors calculated in Fig. \ref{gasmassprofile}, we recalculate the entropy profile in the outskirts as the blue points in the bottom panel of Fig. \ref{T_and_d_profiles}. The clumping corrected entropy profile agrees with the powerlaw relation of \citet{Voit2005} out to our estimate of r$_{200}$ (as was found in \citetalias{Simionescu2011}), but outside r$_{200}$ the entropy profile still flattens and lies below the powerlaw prediction, suggesting that outside the virial radius the entropy profile flattening must be caused by something in addition to clumping. The most likely additional cause for the entropy profile flattening outside r$_{200}$ is that the ICM is out of hydrostatic equilibrium, and that we are seeing cold gas which is still accreting onto the cluster which has yet to be virialised. 

\begin{figure*}
  \begin{center}
    \leavevmode
\hbox{
      \epsfig{figure=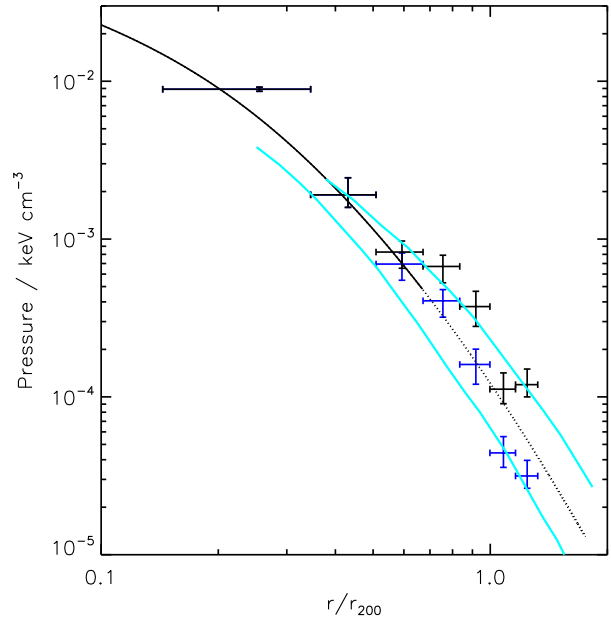, angle=0,
        width=0.5\linewidth}
      }
        
      \caption{Solid line shows the universal pressure profile of \citet{Arnaud2010} out to r$_{500}$, which is extrapolated outwards as the dashed line. Cyan lines shows the range in the universal pressure profile from hydrodynamical simulations of relaxed clusters presented in \citet{Lapi2012}. Black point show the pressures (kn$_{e}$T) using our deprojected densities and deprojected temperatures, which agree with the universal pressure profile within the range of the simulations. Blue points show the pressures corrected by the gas clumping factors found in Fig. \ref{gasmassprofile}, which brings the pressures into better agreement with the universal pressure profile inside r$_{200}$, but which causes them to lie below the universal pressure profile outside r$_{200}$.}
      \label{pressureprofile}
  \end{center}
\end{figure*}

Following the analysis of \citetalias{Simionescu2011} we then compare the pressure profile with the universal pressure profile found in \citet{Arnaud2010} by fitting to clusters explored with \emph{XMM-Newton} out to $\sim$ r$_{500}$ and using simulations beyond this, and see the effect that correcting for clumping has (Fig. \ref{pressureprofile}). As was found in \citetalias{Simionescu2011}, the pressures in the outskirts are systematically higher than the extrapolated universal pressure profile from around r$_{500}$ outwards, but they are still consistent with the universal pressure profile within the range obtained from hydrodynamic simulations \citep{Lapi2012}, shown by the cyan lines in Fig. \ref{pressureprofile}. Taking into account the clumping factor reduces the pressures within r$_{200}$ so that they become more consistent with the universal pressure profile, as was found in \citetalias{Simionescu2011}, but outside r$_{200}$ correcting for clumping causes the pressure to lie below the universal pressure profile. This adds further evidence to the hypothesis that outside r$_{200}$ the gas is out of hydrostatic equilibrium, and that we are seeing cold gas which is accreting onto the cluster. 

It is also possible that the gas mass fraction is being overestimated because we are underestimating the actual mass by using only the thermal pressure in the equation of hydrostatic equilibrium. Numerical simulations (for example \citealt{Lau2009}, \citealt{Burns2010}) indicate that the pressure contribution from random gas motions increases with radius, leading to an underestimate of the true cluster mass if only the thermal pressure is used in the equation of hydrostatic equilibrium. In the bottom panel of Fig. \ref{gasmassprofile} we show the radial profile of the factor by which the mass would need to be underestimated for the cumulative gas fraction not to exceed the mean cosmic baryon fraction, and compare this with the simulated underestimates from \citet{Lau2009} and \citet{Burns2010}. At r$_{200}$ the underestimate needed is $\sim$15 percent, which is highly consistent with the simulated underestimates. Outside the r$_{200}$ the underestimate rises but remains consistent with the simulations of \citet{Lau2009}. 

We therefore conclude that the high gas mass fraction can be explained either by gas clumping or by an understimate of the total mass which is within the expected range from numerical simulations. In reality it is likely that both factors are contributing together, but it is not possible to determine the relative contributions of each individually. 

\begin{table*}
  \begin{center}
  \caption{Comparing our mass determination with previous estimates. The results from \citet{Pointecouteau2005}, \citet{Voigt2006} and \citet{Schmidt2007} were taken from the table compiled in \citet{Comerford2007}.}
  \label{compare_masses}

    \leavevmode
    \begin{tabular}{llll} \hline \hline
    c$_{200}$&M$_{200}$ /10$^{14}$ M$_{\odot}$ &Reference & Comments \\ \hline
        5.3$^{+0.6}_{-0.9}$ & 9.8$^{+1.9}_{-1.0}$ & This work  & Neglects outermost temperature measurement ($\chi^{2}$/d.o.f = 9.0/7)\\ \hline
    5.12$^{+0.40}_{-0.40}$ & 10.0$^{+1.2}_{-1.2}$ & \citet{Pointecouteau2005}& Uses \emph{XMM-Newton} temperature profiles extending out to $\sim$1000 kpc ($\sim$9 arcmins)\\
     5.46$^{+3.22}_{-2.88}$ & 9.7$^{+52.2}_{-8.5}$ & \citet{Voigt2006} & Uses \emph{Chandra} temperature profiles extending out to $\sim$400
kpc ($\sim$3.5 arcmins)\\
     5.86$^{+1.56}_{-1.07}$ & 11.82$^{+4.70}_{-3.55}$ & \citet{Schmidt2007} & Uses \emph{Chandra} temperature profiles extending out to $\sim$400
kpc ($\sim$3.5 arcmins)\\
     - & 10.0$^{+1.2}_{-1.2}$  & \citet{Arnaud2005} & M-T scaling relations (r$_{200}$=1999$^{+77}_{-77}$kpc=17.6$^{+0.7}_{-0.7}$arcmins) \\
     7.5$^{+1.1}_{-0.9}$ & 6.4$^{+0.6}_{-0.6}$ & \citet{George2009} & Underestimated background modelling ($\chi^{2}$/d.o.f = 39/7)\\\hline
     3.45$^{+0.25}_{-0.25}$ & 17.48$^{+1.85}_{-1.85}$ & E11a & Uses \emph{ROSAT} density profile and XMM temperature profile, ($\chi^{2}$/d.o.f = 57/23) \\
     4.89$^{+0.89}_{-0.89}$ & 12.03$^{+3.04}_{-3.04}$ & E11a & Uses \emph{ROSAT} density profile and \emph{BeppoSAX} temperature profile, ($\chi^{2}$/d.o.f = 27/18) \\
     3.52$^{+0.26}_{-0.26}$ & 18.00$^{+2.05}_{-2.05}$ & E11a & Uses \emph{ROSAT} density profile and \emph{Swift} temperature profile, ($\chi^{2}$/d.o.f = 61/19) \\ \hline
    \end{tabular}
  \end{center}
\end{table*}

\section{Comparing \emph{ROSAT} and Suzaku surface brightness profiles.}
\label{sbcomparison}

E11a took the projected densities obtained in G09 and used these to calculate a \textsc{mekal} norm which they then folded through a \emph{ROSAT} PSPC response to obtain the projected surface 
brightness profile that the \emph{Suzaku} results of G09 predict would be obtained using \emph{ROSAT} PSPC. When the surface brightnesses were compared in E11a a significant discrepancy was found outside
14 arcmins, with the \emph{Suzaku} surface brightness being higher than the \emph{ROSAT} at a level of 7.7$\sigma$. 

We repeat the ROSAT imaging analysis of E11a (using the Extended Source Analysis Software of \citealt{Snowden1994}) but instead use a constant threshold count rate for resolving point sources of 0.003 counts per second in the R47 band when using the ftool \textsc{detect} (as used in \citetalias{Eckert2011b}), meaning that the CXB is resolved to the same extent across the entire detector. The PSF of the ROSAT PSPC increases strongly away from the on axis position, meaning that the point source resolving ability is not uniform across the detector, and that it is able to detect fainter point sources nearer to the on-axis position. In order for the CXB to be resolved to the same level uniformly across the entire detector it is necessary to resolve the point sources down to the same threshold flux across the entire detector. 

From Fig. 1. in E11a, which shows the point sources removed in their analysis, it seems from the greater number of resolved point sources closer to the on axis position that a uniform threshold point source flux has not been used. The exact threshold used in removing the point sources is not mentioned in E11a, but we find by repeating their analysis that the mask used can be produced using the ftool \textsc{detect} with a detection threshold of 3 $\sigma$, and that using this reproduced the findings of E11a. The masked PSPC image is shown in Fig. \ref{ROSATimage} along with the circle at the r$_{200}$ value determined in G09, which shows the same point source masking shown in Fig. 1 of E11a. The use of a constant significance means that more point sources have been removed in the cluster emission region (nearer the on-axis position) than in the background region used in E11a, meaning that the CXB has been resolved more in the cluster emission regions. To make this clear, the point sources in Fig. \ref{ROSATimage} are colour coded according to their count rates, showing that the point sources resolved in the cluster region are fainter than can be resolved in the background region. The background level used in E11a is therefore an overestimate. 

\begin{figure}
  \begin{center}
    \leavevmode

      \epsfig{figure=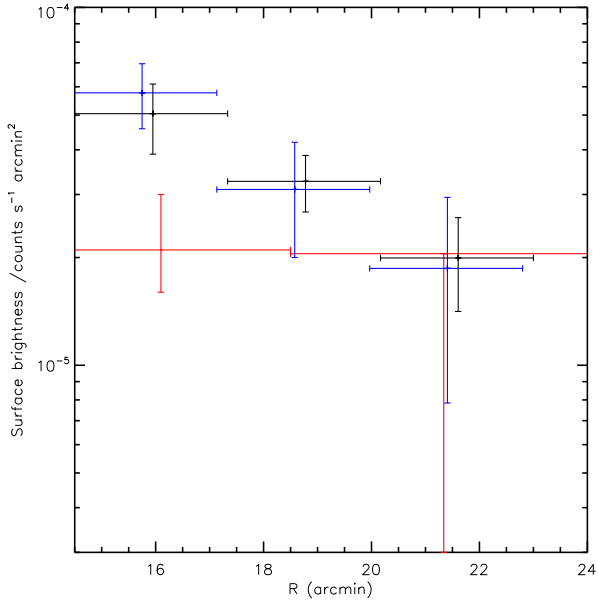, angle=0,
        width=\linewidth}

      \caption{Comparing the projected \emph{ROSAT} PSPC surface brightness profile in the outskirts found using a constant point source threshold count rate of 0.003 counts per second (blue points) with our projected \emph{Suzaku} results folded through the \emph{ROSAT} PSPC response in the R47 band (shown in black). A small horizontal offset has been included between the points to increase the clarity. The errors on the \emph{Suzaku} results add together the one sigma statistical errors of the fits and the one sigma systematic errors of the background modelling. The results from E11a are shown in red, where the surface brightness in their outer bin is a 90 percent upper limit.} 
      \label{sb_profile_fold}
  \end{center}
\end{figure}

 When we use a constant threshold count rate of 0.003 counts per second in the R47 band to resolve point sources, the resulting surface brightness profile in the outskirts, extracted in the same regions as our Suzaku observations, is shown as the blue points in Fig. \ref{sb_profile_fold}. We then use our best fit projected \emph{Suzaku} results to produce the projected surface brightness profile we would expect to see with \emph{ROSAT} in the R47 band. This is shown as the black points in Fig. \ref{sb_profile_fold}, where the statistical error on the fits and the systematic error in the background modelling have been added to produce the error bars shown. We find complete agreement between the ROSAT results and the Suzaku results. 

The dominant cause of the discrepancy between the \emph{Suzaku} results of G09 and the
\emph{ROSAT} results of E11a is the underestimate of the background level in G09, as
when a constant flux threshold
is used for the \emph{ROSAT} analysis the predicted surface brightness from the G09
results is still higher than the \emph{ROSAT} measurement at the 5$\sigma$ level in the outskirts.

The results in E11a are shown as the red points (which used the same binning as G09) in Fig. \ref{sb_profile_fold}, and we see that the use of a constant significance in removing point sources (rather than a constant flux) in E11a has caused the CXB to be resolved more near the centre of the pointing than in the background regions, thus causing the background to be overestimated and underestimating the surface brightness in the outskirts of the cluster. This subsequently means that the densities found in the cluster outskirts in E11a (which are shown in Fig. \ref{T_and_d_profiles}) are also likely to be underestimated.

E11a used the density profile obtained with \emph{ROSAT} and the temperature profiles obtained with \emph{XMM-Newton}, \emph{BeppoSAX} and \emph{Swift} to perform a similar mass analysis to the one we present and their results are tabulated for comparison in table \ref{compare_masses}. Whereas the masses using the \emph{BeppoSAX} temperatures and the \emph{ROSAT} densities are consistent with our results, those obtained using the \emph{XMM-Newton} temperatures and \emph{Swift} temperatures are higher than our results. The ROSAT mass analysis of E11a used the density profile out to 17 arcmins,
which from Fig. \ref{T_and_d_profiles} is completely consistent with our \emph{Suzaku} density
profile, so the differences in
the mass determinations are due to the differences in the temperature profiles,
as is also described in Appendix A of E11a.

\section{Summary}
We have used new \emph{Suzaku} observations of the background regions surrounding PKS 0745-191 to perform a thorough analysis of the error in the background modelling performed, allowing a more accurate measurement of the thermodynamic properties of the ICM out to 23$'$, beyond which we do not detect any statistically significant emission. We have highlighted and corrected the problems with the background modelling used in G09. 

We find a temperature profile which is consistent with those already obtained using \emph{XMM-Newton} and \emph{Swift}, and conclude that the low temperatures found in G09 in the 6.0$'$-14.5$'$ region were the result of incorrect background modelling. Our higher temperatures cause the entropy profile to obey the powerlaw relation, S $\propto$ r$^{1.1}$, to higher radius than in G09 before flattening and decreasing above 15$'$ ($\approx$ 1.7 Mpc). This brings the average temperature scaled entropy profile into strong agreement with those from other clusters investigated with Suzaku out to r$_{200}$.  

The higher temperatures found in the 6.0$'$-14.5$'$ region improve the fitting statistics for modelling the ICM as being in hydrostatic equilibrium in an NFW potential from $\chi^{2}$/d.o.f =39/7 obtained in G09 to $\chi^{2}$/d.o.f =9.0/7 using all of the temperature profile except the outer annulus. The fit to an NFW profile is much poorer if the outermost temperature annulus is included, further indicating that beyond $\sim$17$'$ the ICM is out of hydrostatic equilibrium. The estimates of M$_{200}$, c$_{200}$ and r$_{200}$ are consistent with previous results and with the scaling relations of \citet{Arnaud2005}, as shown in table \ref{compare_masses}, thus resolving the problem with the G09 measurements which found unusually small M$_{200}$ and  r$_{200}$ values.   
We find that r$_{200}$ lies in the range 16.7-18.4 arcmins (1.9-2.1 Mpc). This means that we have probed the ICM of PKS 0745-191 to at least 1.25r$_{200}$ = 23$'$, beyond which we find no statistically significant emission. 

We find that the gas mass fraction increases significantly above the mean cosmic baryon fraction from around r$_{200}$ onwards, and is consistent with the profile found in \citetalias{Simionescu2011} for the Perseus cluster. This suggests that the ICM may be clumped in the outskirts, causing the gas density to be overestimated. Correcting for the clumping factor (if we assume that the true gas mass fraction does not increase above the mean cosmic baryon fraction purely because of gas clumping) causes the entropy profile to obey the powerlaw relation S $\propto$ r$^{1.1}$ out to our estimate of r$_{200}$ (as was observed in \citetalias{Simionescu2011} for the Perseus cluster) but above r$_{200}$ the entropy profile still flattens and lies below the powerlaw prediction, suggesting that something other than clumping is responsible for the entropy profile flattening there. The most likely cause is that outside r$_{200}$ the ICM is out of hydrostatic equilibrium and that we as seeing cold gas which has not yet been virialised as it accretes onto the cluster. This hypothesis is further supported by comparing the clumping corrected pressure profile with the universal pressure profile of \citet{Arnaud2010}. Between r$_{500}$ and r$_{200}$ the clumping correction reduces the pressure and improves the agreement with the universal pressure profile, as was found in \citetalias{Simionescu2011}. However outside r$_{200}$ the clumping correction causes the gas pressure to lie below the extrapolated universal pressure profile, suggesting that outside r$_{200}$ we are seeing cold gas which is accreting onto the cluster. 

It is also possible that the observed gas mass fraction exceeds the cosmic mean baryon fraction in the outskirts because the total mass is being underestimated by using only the thermal pressure in the equation of hydrostatic equilibrium. The total mass underestimate needed for the cumulative gas mass fraction to stay below the cosmic mean baryon fraction agrees with the expected underestimate of the total mass from numerical simulations \citep{Lau2009}, which is caused by the increasing contribution of non-thermal pressure in the outskirts which cannot be directly measured at present. In reality it is likely that a combination of gas clumping and non-thermal pressure support is responsible for the high gas mass fraction and the flattening of the entropy profile in the outskirts. 

Our results indicate that a constant threshold flux was not used for resolving point sources in ROSAT PSPC analysis of E11a, which has caused greater point source extraction nearer to the on-axis position due to the increase in the PSF (and the decrease in the point source detection sensitivity) away from the centre of the PSPC detector. This means that the CXB was not resolved uniformly across the PSPC pointing, with the threshold flux for the point sources removed in the background regions being higher than it was for the cluster emission regions. This has caused the background to be overestimated, thus underestimating the surface brightness and density in the cluster outskirts.

We predict from our \emph{Suzaku} spectral fits a projected \emph{ROSAT} surface brightness profile which is consistent with that obtained with \emph{ROSAT} when the errors in the background modelling are taken into account and when a constant point source threshold is used for removing point sources in the \emph{ROSAT} PSPC image.

Using our background pointings for PKS 0745-191 (which lies 3 degrees above the galactic midplane), we find no statistically significant evidence for the presence of the absorbed bremsstrahlung background component (due to unresolved dM stars) found in \citet{Masui2009} in their study of a background pointing in the galactic midplane. This supports the hypothesis that this background component observed in \citet{Masui2009} originates from unresolved dM stars whose number density decreases rapidly away from the galactic midplane (as shown in Fig. 4 of \citealt{Masui2009}).

\label{summary}

\section*{Acknowledgements}

SAW is supported by STFC, and ACF thanks the Royal Society. MRG acknowledges
support from an NSF Graduate Research Fellowship and NASA grant NNX10AR49G. This research has used data from the $Suzaku$
telescope, a joint mission between JAXA and NASA.

\bibliographystyle{mn2e}
\bibliography{PKS0745-191_paper}

\appendix
\section[]{}
\label{sec:appendix}

\begin{figure*}
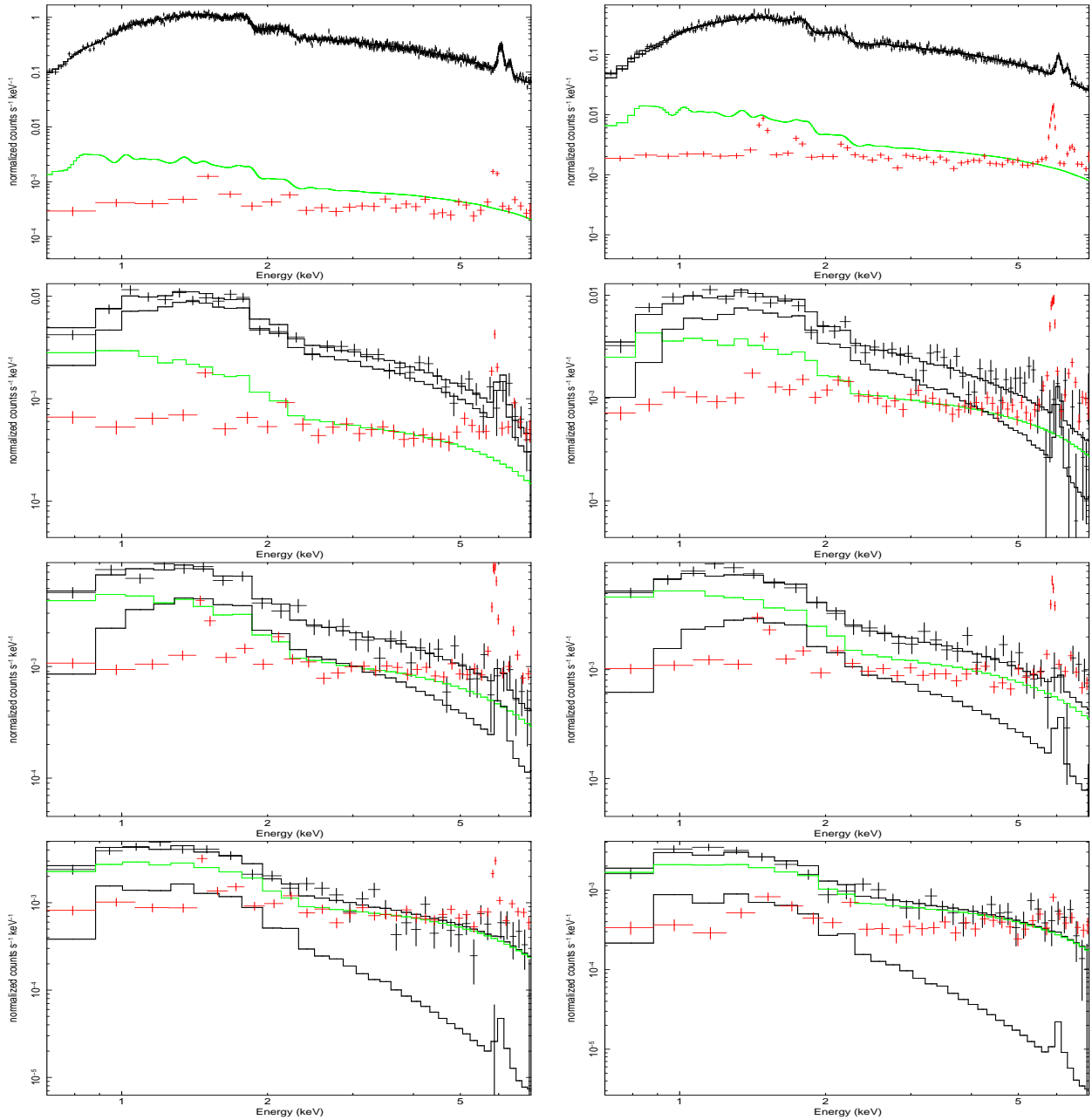

  \begin{center}
     \hbox{\epsfig{figure=FigA1a.ps, height=1.0\columnwidth,width=0.5\columnwidth, angle=-90 }
     \epsfig{figure=FigA1b.ps, height=1.0\columnwidth,width=0.5\columnwidth, angle=-90 }
     }
      \hbox{\epsfig{figure=FigA1c.ps, height=1.0\columnwidth,width=0.5\columnwidth, angle=-90 }
     \epsfig{figure=FigA1d.ps, height=1.0\columnwidth,width=0.5\columnwidth, angle=-90 }
     }    
          \hbox{\epsfig{figure=FigA1e.ps, height=1.0\columnwidth,width=0.5\columnwidth, angle=-90 }
     \epsfig{figure=FigA1f.ps, height=1.0\columnwidth,width=0.5\columnwidth, angle=-90 }
     }   
            \hbox{\epsfig{figure=FigA1g.ps, height=1.0\columnwidth,width=0.5\columnwidth, angle=-90 }
     \epsfig{figure=FigA1h.ps, height=1.0\columnwidth,width=0.5\columnwidth, angle=-90 }
     } 
     
      \caption{Spectral fitting for the cluster emission. In reading order we show the azimuthally averaged spectra used to generate the results in Fig. \ref{T_and_d_profiles}: the corresponding annuli are 0.0$'$-2.5$'$, 2.5$'$-6.0$'$, 6.0$'$-8.8$'$, 8.8$'$-11.7$'$, 11.7$'$-14.5$'$, 14.5$'$-17.3$'$, 17.3$'$-20.2$'$, 20.2$'$-23$'$. The data from each detector and each editing mode were fitted simultaneously and are added here for display purposes only. The black lines through the points represent the best projected fits (background plus cluster emission) which are shown for simplicity, while the lower black line shows the cluster emission. The green line shows the X-ray background level, and the red points show the NXB level which is subtracted from the X-ray spectra. The fit quality is the same for the deprojected fits but we show the projected fits for simplicity.}
      \label{cluster_spectra}
  \end{center}
\end{figure*}

\begin{figure*}
  \begin{center}
     \hbox{\epsfig{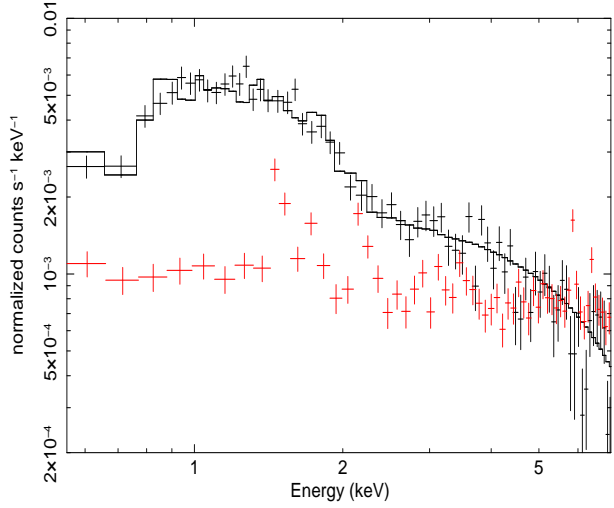}
     }

   \caption{Spectral fitting to all of the \emph{Suzaku} background regions between 23$'$-41$'$ simultaneously. The spectrum is well fit using two \textsc{apec} components of solar metallicity and zero redshift to model the galactic emission; an unabsorbed component at 0.1 keV modelling the LHB and an absorbed component at 0.6 keV. An absorbed powerlaw of index 1.4 is used to model the CXB. The NXB level which is subtracted from the X-ray spectra is shown overplotted as the red points. }
      \label{bkg_spectra}
  \end{center}
\end{figure*}



\label{sec:appendixROSAT}
\begin{figure*}
  \begin{center}
     \hbox{\epsfig{figure=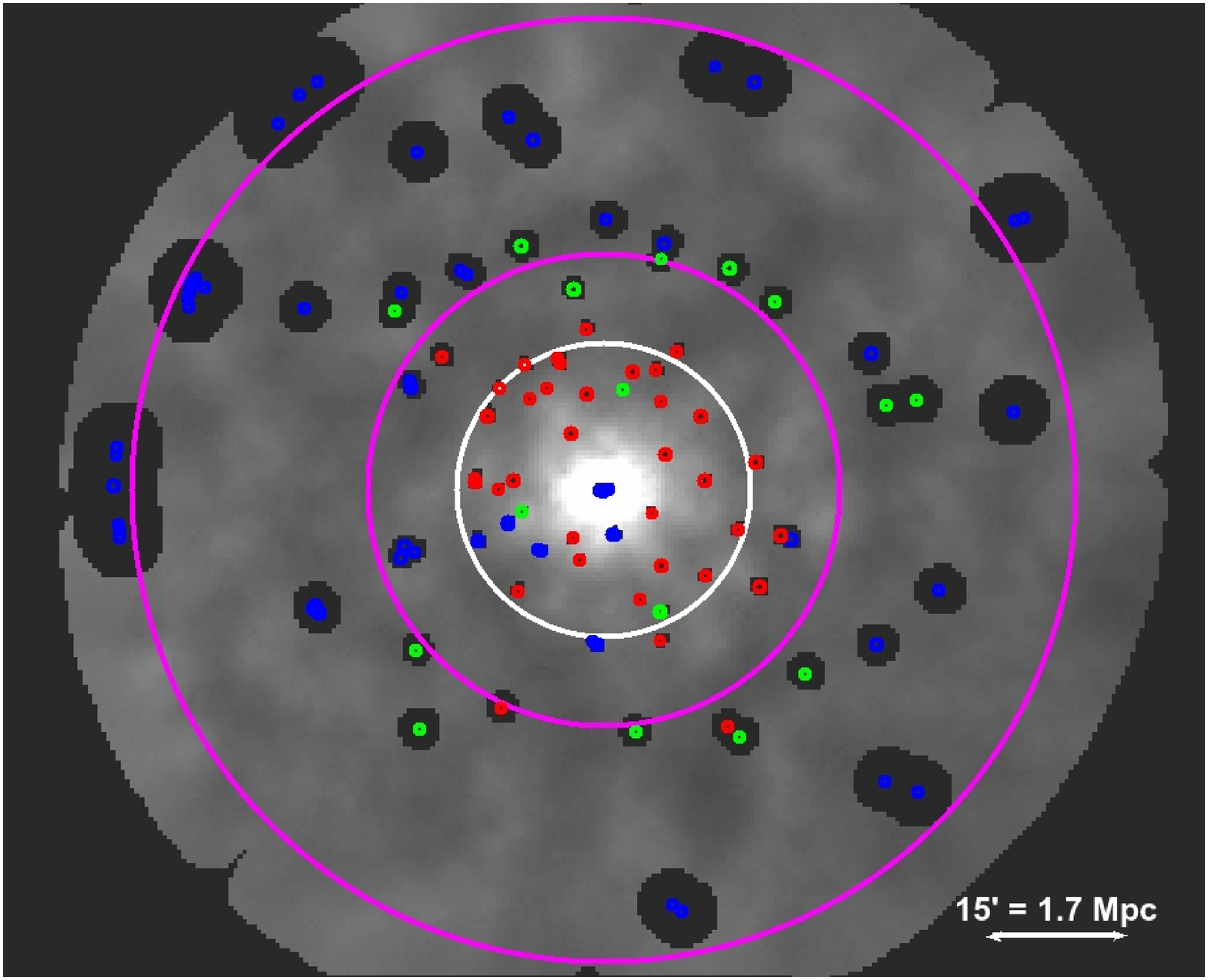, height=1.5\columnwidth}
     }

   \caption{Background subtracted, exposure corrected and smoothed ROSAT PSPC image in the R47 band following the procedure of E11a. The point source removal is exactly the same as that used in E11a (compare the inner part with Fig. 1 in E11a), and we find that this is achieved using a constant significance of 3 $\sigma$ in the ftool \textsc{detect} when removing point sources. Due to the increase in the PSF with off axis angle, this means that the CXB has been resolved more deeply closer to the on-axis position, leading to an overestimate of the background in E11a. The colours of the point sources correspond to their count rates; red point sources have count rates less than 0.002 cts/s, the greens ones are between 0.002-0.003 cts/s, and the blue ones are $>$0.003 cts/s. The white circle at the r$_{200}$ value determined in G09 is the same as that shown in E11a in their Fig. 1 and allows a direct comparison. The background region used in E11a is the pink annulus between 25$'$ and 50$'$.    }
      \label{ROSATimage}
  \end{center}
\end{figure*}

\end{document}